\documentclass[graybox, secnum]{svmult}

\usepackage{mathptmx}       
\usepackage{helvet}         
\usepackage{courier}        
\usepackage{type1cm}        
\usepackage{makeidx}         
\usepackage{graphicx}        
\usepackage{multicol}        
\usepackage[bottom]{footmisc}
\usepackage{hyperref}        
\usepackage{soul}            
\hypersetup{colorlinks=true,urlcolor=blue}
\usepackage[square,numbers]{natbib}
\usepackage{amsmath,amsfonts}
\usepackage{shuffle}
\makeindex             

\def\ap{{\alpha^{\prime}}}
\def\halfap#1{\Big({\ap\over 2}\Big)^{\mkern-4mu #1}}
\def\a{\alpha}
\def\b{{\beta}}
\def\g{{\gamma}}
\def\d{{\delta}}
\def\e{{\epsilon}}
\def\l{\lambda}
\def\k{{\kappa}}
\def\s{{\sigma}}
\def\t{{\theta}}
\def\om{{\omega}}
\def\lb{{\overline\lambda}}
\def\llb{(\l\lb)}
\def\wb{{\overline w}}
\def\half{{1\over 2}}
\def\p{{\partial}}
\def\pb{{\overline\partial}}

\def\bar{\overline}
\def\({\left(}
\def\){\right)}

\def\cF{{\cal F}}

\def\cK{{\cal K}}
\def\cI{{\cal I}}

\def\cZ{{\cal Z}}

\def\eqref#1{(\ref{#1})}
\def\beq{\begin{equation}}
\def\eeq{\end{equation}}

\def\bA{\mathbb{A}}

\def\AYM{A^{\rm SYM}}

\def\perm#1{{\rm perm}#1}

\def\ImOmega{\Im\Omega}
\def\Im{\mathop{{\rm Im}}} 
\def\halfap#1{\Big({\ap\over 2}\Big)^{\mkern-4mu #1}}

\allowdisplaybreaks

\begin{document}
\title*{Pure spinor formulation of the superstring and its applications}
\author{Nathan Berkovits and Carlos R. Mafra}
\institute{Nathan Berkovits \at
{\sl ICTP South American Institute for Fundamental Research,
Instituto de F\' isica Te\' orica, UNESP -- Universidade Estadual Paulista,
Rua Dr. Bento T. Ferraz 271, 01140-070, S\~ ao Paulo, SP, Brasil}
\email{nathan.berkovits@unesp.br}
\and Carlos R. Mafra \at {\sl STAG Research Centre and Mathematical Sciences, University of Southampton, Highfield, Southampton SO17 1BJ, UK} \email{c.r.mafra@soton.ac.uk}}
%
%
\maketitle
\abstract{The pure spinor formalism for the superstring has the advantage over
the more conventional Ramond-Neveu-Schwarz formalism of being manifestly
spacetime supersymmetric, which simplifies the computation of multiparticle
and multiloop amplitudes and allows the description of Ramond-Ramond
backgrounds. In addition to the worldsheet variables of the
Green-Schwarz-Siegel action, the pure spinor formalism includes bosonic ghost
variables which are constrained spacetime spinors and are needed for covariant
quantization using a nilpotent BRST operator.\vskip5pt
In this review, several
applications of the formalism are described including the explicit computation
in D=10 superspace of the general disk amplitude with an arbitrary number of
external massless states, genus one amplitudes with up to seven external
states, genus two amplitudes with up to five external states, and the
low-energy limit of the genus three
amplitude with up to four external states. The pure spinor
formalism has also been used to covariantly quantize the superstring in an
$AdS_5\times S^5$ background and might be useful for proving the AdS-CFT
correspondence in the limit of small AdS radius. \vskip5pt
This is an overview written for the ``Handbook of Quantum Gravity'', eds. C. Bambi, L. Modesto and I. Shapiro.}

\section*{Keywords} 
superstring, supersymmetry, amplitudes, pure spinors.

\section{Introduction}

At the present time, superstring theory is the only formalism available for computing perturbative scattering amplitudes of gravitons without ultraviolet quantum-mechanical
divergences. Although comparing these scattering amplitudes with experiments is unlikely in the near future, various properties of these amplitudes such as spacetime
supersymmetry and duality symmetry might have testable low-energy implications.

Using the conventional Ramond-Neveu-Schwarz (RNS) formalism of the superstring, the complicated nature of vertex operators for spacetime fermions and the need to sum over spin
structures has made it difficult to compute amplitudes involving external fermions or to compute multiloop amplitudes. Furthermore, backgrounds involving Ramond-Ramond fields necessary
for the AdS-CFT correspondence are difficult to describe in the RNS formalism. 

In 2000, a new formalism for the superstring was constructed in which spacetime supersymmetry is manifest and there is no need to sum over spin structures \cite{psf}. In addition to the worldsheet
variables $(x^m, \theta^\alpha)$ of the Green-Schwarz formalism \cite{GS} for $m=0$ to 9 and $\a=1$ to 16, this new formalism includes the fermionic momenta variables $d_\alpha$ of Siegel \cite{siegel} as well as bosonic ghost variables $(\lambda^\alpha, w_\alpha)$ constrained to satisfy $\lambda \gamma^m \lambda=0$. This constraint implies that $\lambda^\alpha$ is a $D=10$ ``pure 
spinor" as defined by Cartan with 11 independent components, and the conformal anomaly contribution of $+22$ from $(\lambda^\alpha, w_\alpha)$ cancels the conformal
anomaly contribution of $+10-32=-22$ from $x^m$ and $(\theta^\alpha, d_\alpha)$. Generalizing a supersymmetric field theory observation of Howe \cite{Howe1, Howe2}, physical superstring states
in this ``pure spinor formalism" are defined using the nilpotent BRST operator $Q = \oint dz \lambda^\alpha d_\alpha$ and, unlike in the Green-Schwarz formalism, covariant
quantization is straightforward.\footnote{Other early
applications of D=10 pure spinors include \cite{Nilsson:1985cm} in off-shell
super-Yang-Mills and \cite{Hughston:1987km} in classical superstrings.} A similar BRST operator is useful for describing $d=11$ supergravity \cite {sugra1, Mtheory1, sugra2}, and more details on pure spinor applications in supersymmetric field theories can be found in the review of Martin Cederwall \cite{cederwall}.

Over the last 20 years, this pure spinor formalism has been used to compute various multiparticle and multiloop amplitudes in superstring theory including several amplitudes which have not yet been computed
using the RNS formalism. All amplitudes computed using both the RNS and pure spinor formalisms have been shown to coincide, however, the pure spinor computations are typically much
more efficient since there is no sum over spin structures and amplitudes are automatically expressed in $D=10$ superspace. Nevertheless, a proof of equivalence of the RNS and pure
spinor formalisms for the superstring is still lacking. 

A promising approach towards proving this equivalence involves a recently constructed formalism for the superstring which includes $\theta^\alpha$
and an unconstrained bosonic spacetime spinor worldsheet variable $\Lambda^\alpha$, in addition to the usual N=1 worldsheet supersymmetric RNS matter and ghost variables. \cite{BRNSGSS1, BRNSGSS2} This new formalism, named the B-RNS-GSS formalism since it combines features of the RNS, Green-Schwarz-Siegel and pure spinor formalisms, is both N=1 worldsheet supersymmetric and D=10
spacetime supersymmetric and acts as a bridge between the RNS and pure spinor formalisms. It can be related to the RNS formalism by treating $(\theta^\alpha, \Lambda^\alpha)$
as non-minimal variables which decouple from the BRST cohomology, and can be related to the pure spinor formalism by ``twisting" the N=1 superconformal generators into N=2 superconformal generators so that all worldsheet variables carry integer conformal weight. Work is in progress on computing scattering amplitudes using the B-RNS-GSS formalism and proving that the amplitudes coincide with those computed using the RNS and pure spinor formalisms. Since multiloop amplitude computations using the RNS and pure spinor formalism have different
types of subtleties, it is expected that the B-RNS-GSS formalism will be useful for relating these subtleties.

Just as the RNS formalism for the superstring can be described in any curved background which preserves N=1 worldsheet supersymmetry, the pure spinor formalism can be described in any curved background in which the BRST current $\lambda^\alpha d_\alpha$ remains nilpotent and holomorphic \cite{BerkovitsHowe}. This allows not only the Calabi-Yau backgrounds which can be described using the RNS formalism, but also
any curved supergravity background in which the D=10 supergravity equations of motion are satisfied to lowest order in $\alpha'$. 
For example, unlike the RNS formalism, the pure spinor formalism can be used to covariantly quantize the superstring in an $AdS_5\times S^5$ 
Ramond-Ramond background which is dual to ${\cal N}=4$ $D=4$ super-Yang-Mills through the AdS-CFT correspondence.

Although this important application will not be discussed in later sections of
the review, quantum consistency of the $AdS_5\times S^5$ background has been
proven \cite{quantum} using the pure spinor formalism. To prove quantum consistency to all orders in
$\alpha'$, it was shown using symmetry arguments that any potential BRST
anomalies coming from quantum corrections can be cancelled by the addition of
local counterterms to the worldsheet action. It was also shown using BRST
arguments that the classical non-local conserved currents related to
integrability can be extended to quantum non-local conserved currents.

The construction of BRST-invariant vertex operators for half-BPS states in an
$AdS_5\times S^5$ background was recently achieved \cite{BerkFleury1,
BerkFleury2, BerkFleury3}, and work is in progress on using these vertex 
operators for the computation of scattering amplitudes. The structure of the
vertex operators and the pure spinor worldsheet action in an $AdS_5\times S^5$
background is more complicated than in a flat background, however, the
manifest $PSU(2,2|4)$ isometry of the construction should be useful in
simplifying the amplitude computations. An important open question is how to
generalize the super-Poincar\'e invariant BRST cohomology methods which are
described in this review to BRST cohomology methods with $PSU(2,2|4)$
invariance.

In the limit of small AdS radius, the pure spinor version of the $AdS_5\times S^5$ worldsheet action has been shown to reduce to a BRST-trivial topological
action plus a small $PSU(2,2|4)$-invariant deformation term \cite{topological1, topological2}. In this limit, the dual theory is ${\cal N}=4$ $D=4$ super-Yang-Mills at weak coupling, and it has been conjectured that the topological action describes
free super-Yang-Mills and the deformation describes the cubic super-Yang-Mills interaction term. The topological action and deformation term are constructed by combining the $x^m$
and $\lambda^\alpha$ bosonic worldsheet variables of the pure spinor formalism into a twistor-like variable which transforms linearly under the $SO(4,2)\times SO(6)$ bosonic subgroup of $PSU(2,2|4)$. Similar twistor variables have been extremely useful for computing perturbative scattering amplitudes of ${\cal N}=4$ $D=4$ super-Yang-Mills \cite{witten}, and it would not be surprising
if the two types of twistor variables are related through the AdS-CFT correspondence.

If this conjecture could be verified, it would provide a proof of the AdS-CFT correspondence in the case of $AdS_5\times S^5$. A proof of the AdS-CFT correspondence in the simpler case of $AdS_3\times S^3$ was established by Eberhardt, Gaberdiel and Gopakumar in \cite{ads3} using a ``hybrid" formalism of the superstring which can be interpreted as a six-dimensional version of the $D=10$ pure spinor formalism. It is very suggestive that twistor-like variables were used in their proof, and that Gaberdiel and Gopakumar were recently able to generalize their twistor-like construction of the spectrum of  $AdS_3\times S^3$ at zero radius to the more interesting case of $AdS_5\times S^5$ at zero radius \cite{gabardiel}.

After a brief review of the pure spinor formalism and the superspace
formulation of ten-dimensional super Yang-Mills theory in sections 2.1 and
2.2, section 2.3 will showcase its applications to the computation of
scattering amplitudes in a flat background. From the complete genus-zero
amplitudes with an arbitrary number of external massless states to the
low-energy limit of the massless four-point amplitude at genus three, the pure
spinor formalism and related techniques played a crucial role in determining
their manifestly supersymmetric forms. Finally, section 2.4 will discuss how
these amplitudes have been used to test S-duality conjectures.

\section{The pure spinor formalism and scattering amplitudes}

\subsection{Ten-dimensional super-Yang-Mills theory in superspace}

There is a super-Poincar\'e description of $D=10$ super-Yang-Mills (SYM) in superspace
\cite{siegelSYM,wittentwistor}
that describes the gluon and gluino states via Lie algebra-valued superfield connections
$\mathbb{A}_\a(x,\t)$ and $\mathbb{A}_m(x,\t)$ satisfying the non-linear constraint
$\{\nabla_\a,\nabla_\b\}=\gamma^m_{\a\b}\nabla_m$, where $\nabla_\a= D_\a - \bA_\a$ and
$\nabla_m = \p_m - \bA_m$ are supercovariant derivatives and
\beq\label{covD}
D_\a={\p\over\p\t^\a}+\half (\g^m\t)_\a\p_m
\eeq
is the superspace derivative satisfying $\{D_\a,D_\b\}=\g^m_{\a\b}\p_m$. The ten-dimensional
superspace coordinates $(x,\t)$ are composed of a $SO(9,1)$ Lorentz vector $x^m$, where $m=1, \ldots,10$, and a Weyl spinor
$\t^\a$, where $\a=1, \ldots,16$. In ten dimensions, the Lorentz group has two inequivalent
spinor representations, denoted Weyl and anti-Weyl. They are
distinguished by the position of the spinor index, upstairs for Weyl $\Psi^\a$ and downstairs for
anti-Weyl $\chi_\a$ which cannot be raised or lowered.
The gamma matrices $\g^m_{\a\b}$ and $\g_m^{\a\b}$ are the $16\times16$ off-diagonal symmetric Pauli
matrices of the $32\times32$ Dirac matrices $\Gamma^m$ of the $SO(9,1)$ Clifford algebra
$\{\Gamma^m,\Gamma^n\}=2\eta^{mn}\mathbb{I}_{32\times32}$. They satisfy
$\g^m_{\a\b}(\g^n)^{\b\rho}+\g^n_{\a\b}(\g^m)^{\b\rho}=2\eta^{mn}\d_\a^\rho$.

The non-linear equations of motion following from the above constraint have
linearized counterparts written in terms of linearized superfield connections
$A_\a(x,\t)$, $A^m(x,\t)$ and
their field-strengths $W^\a(x,\t)$, and $F^{mn}(x,\t)$,
\begin{align}
\label{singleS}
D_{\a} A_{\b} + D_\b A_\a & = \g^m_{\a\b} A_m\,, &D_\a A_m &= (\g_m W)_\a + \p_m A_\a  \cr
D_\a F_{mn} & = \p_{[m} (\g_{n]} W)_\a\,,
&D_\a W^{\b} &= {1\over 4}(\g^{mn})^{\phantom{m}\b}_\a F_{mn}\,.
\end{align}
These linearized superfields will enter the expressions for the massless vertex operators of
the pure spinor formalism and will be the main actors in the composition of {\it pure spinor
superspace} expressions to be reviewed below. In this context, it is essential to know how
these superfields are expanded in a series of $\t$ variables.

The linearized superfields can be expanded in the so-called Harnad-Shnider gauge $\t^\a
A_\a(x,\t)=0$ in terms of the gluon $e^m_i$ and gluino $\chi^\a_i$ polarizations of a particle
state labelled by $i$ \cite{HSgauge}. For convenience we strip off
the universal plane-wave factor $e^{k\cdot x}$
that carries all the $x$ dependence from the superfields and define their $\t$-dependent factor as
$A^i_\a(x,\t) = A^i_\a(\t)e^{k\cdot x}$ etc. One can show that
\begin{align}
A^i_{\alpha}(\theta)&=
{1\over 2}(\theta\gamma_m)_\alpha e^m_i
+{1\over 3}(\theta\gamma_m)_\alpha (\theta\gamma^m \chi_i)
- {1\over32}(\theta\gamma_m)^\alpha(\theta\gamma^{mnp}\theta)f^i_{np} \label{lintheta}\\
&+ {1\over60}(\t\g_m)_\alpha(\t\g^{mnp}\t) k^i_n (\chi^i \g_p\t)
+ {1\over1152}(\t\g_m)_\a(\t\g^{mnp}\t)(\t\g^{pqr}\t) k_i^n f_{i}^{  qr}  + \ldots \notag\\
A_i^m(\t)&= e_i^m
+(\t\g^m \chi_i)
-{1\over8}(\t \g^{mpq}\t) f_i^{pq}
+{1\over12}(\t\g^{mnp}\t) k_i^n (\chi_{i} \g^{p}\t)\notag\\
&+{1\over 192}(\t\g^{m}_{\phantom{m}nr}\t)(\t\g^{r}_{\phantom{r}pq}\t) k_i^n f_i^{  pq}
-{1\over480} (\t\g^{m}_{\phantom{m}nr}\t)(\t\g_{\phantom{r}pq}^{r}\t)k_i^n k_i^p (\chi_{i}\g^{q}\t)  + \ldots
\notag\\
W_i^\a(\t)&=
\chi^\alpha_i
+{1\over 4}(\t \g^{mn})^{\alpha} f^i_{mn}
-{1\over 4}(\t \g_{mn})^{\alpha} k_i^m (\chi_i \g^{n}\t)
-{1\over48} (\t\g^{\phantom{m}q}_{m})^\alpha(\t\g_{qnp}\t) k_i^m f^{np}_i \notag\\
&+{1\over96}  (\t\g^{\phantom{m}q}_{m})^{\alpha}(\t\g_{qnp}\t) k_i^m k_i^n (\chi_{i}\g^{p}\t)
- {1\over 1920}(\t\g^{\phantom{m}r}_{m})^{\alpha}(\t\g^{\phantom{nr}s}_{nr}\t)(\t\g_{spq}\t) k_i^m k_i^n f^{ pq}_{i} 
 + \ldots  \notag\\
F_i^{mn}(\t)&=
f_i^{mn}
- k_i^{[m}(\chi_i\g^{n]}\t)+{1\over 8}(\t\g_{pq}^{\phantom{pq}[m}\t) k_i^{n]} f_i^{ pq}
 -{1\over12}(\t\g^{\phantom{pq}[m}_{pq}\t) k_i^{n]} k_i^p (\chi_{i}\g^{q}\t)\notag\\
& -{1\over192}(\t\g^{\phantom{ps}[m}_{ps}\t) k_i^{n]} k_i^p f^{ qr}_i(\t\g^{s}_{\phantom{s}qr}\t)
+ {1\over 480}(\t\g^{[m}_{\phantom{[m}ps}\t) k_i^{n]} k_i^p k_i^q (\chi_{i}\g^{r}\t)(\t\g^{s}_{\phantom{s}qr}\t) + \ldots
\, , \notag
\end{align}
where $f^{mn}_i = k^m e^n_i - k^n e^m_i$ is the linearized field strength of the $i$th gluon and
the terms in the ellipsis of order $\t^{>5}$ will not contribute in pure spinor superspace
expressions to be reviewed below.

\subsection{Non-minimal pure spinor formalism}

It is customary to distinguish two very closely related versions of the pure spinor
formalism:
minimal \cite{MPS} and non-minimal \cite{NMPS}. Both are based on the ideas of \cite{psf} but the non-minimal incarnation
introduces new variables on the worldsheet and admits a simpler ``topological'' multiloop amplitude
prescription.

The left-moving sector of the
non-minimal pure spinor formalism is composed of the fields $\p x^m, p_\a, w_\a, s^\a$ of
conformal weight one and of $\t^\a,\l^\a, \lb_\a, r_\a$ of conformal weight zero,
where $m=0,1 \ldots,9$ and $\a=1, \ldots,16$ are the vector and spinorial indices of $SO(10)$.
The world-sheet action is
\beq\label{NMPSaction}
S = {1\over 2\pi \ap}\int d^2z \( \p x^m \pb x_m + \ap p_\a \pb \t^\a
- \ap w_\a \pb\l^\a
- \ap \wb^\a \pb\lb_\a + \ap s^\a\pb r_\a \),
\eeq
and $\ap$ denotes the inverse string tension.
The field $\l^\a$ is bosonic and satisfies the {\it pure
spinor constraint}
\beq\label{PSconstraints}
\l^\a\g^m_{\a\b}\l^\b = 0\,.
\eeq
The field $\lb_\a$ is bosonic while $r_\a$ is fermionic and they satisfy the constraints
\beq\label{nonminc}
\lb_\a\g_m^{\a\b}\lb_\b = 0, \quad \lb_\a\g_m^{\a\b} r_\b = 0\,.
\eeq
The OPEs of the matter variables are given by
\beq\label{matterOPE}
X^m(z,\bar z)X_n(w,\bar w) \sim -{\ap\over2}\d^m_n\ln|z-w|^2,\qquad
p_\a(z)\t^\b(w) \sim {\d_\a^\b\over z-w}\,,
\eeq
while the OPEs of the ghost variables do not follow from a free-field calculation
due to the constraints above. In certain circumstances, however, the variables $(w_\a,\l^\a)$ can be viewed
as a conjugate pair with canonical OPE.
The Green--Schwarz constraint $d_\a(z)$, the supersymmetric momentum $\Pi^m(z)$
have conformal weight $+1$ and are given by
\begin{align}
\label{OPEs}
d_\a(z) &= p_\a - {1\over \ap}(\g^m\t)_\a \p x_m - {1\over 4\ap}(\g^m \t)_\a(\t \g_m \p\t),\\
\Pi^m(z) &= \p x^m + \half (\t\g^m \p\t)\,.\notag
\end{align}
These fields satisfy the following OPEs
\begin{align}
d_\a(z)d_\b(w) &\sim -  {\g^m_{\a\b}\Pi_m\over z- w},\qquad
\Pi^m(z)\Pi^n(w) \sim -  {\eta^{mn}\over (z - w)^2},\\
d_\a(z) \Pi^m(w) &\sim  {(\g^m\p\t)_\a\over z- w},\qquad
\notag
\end{align}
In addition, if $f(x(w),\t(w))$ does not depend on derivatives $\p^k x$ nor $\p^k \t$ 
\beq
\label{dK}
d_\a(z) f(x(w),\t(w)) \sim  {D_\a f\over z- w}, \qquad
\Pi^m(z) f(x(w),\t(w)) \sim -  {k^m f\over z- w}
\eeq
The non-minimal BRST charge
\beq\label{BRST}
Q=\oint (\l^\a d_\a + \wb^{\a}r_{\a}),
\eeq
can be shown to be nilpotent $Q^2=0$ using the OPEs \eqref{OPEs} and
the pure spinor constraint \eqref{PSconstraints}. Physical states are required to be in the
cohomology of \eqref{BRST} and it will be shown below that the cohomology is independent on 
the quartet of non-minimal variables
$(\bar w^\a,\lb_\a,s^\a,r_\a)$.

The constraint \eqref{PSconstraints} implies that the conjugate momentum $w_\a$ to the pure spinor
$\l^\a$ can only appear in gauge-invariant combinations under
\beq\label{varw}
\d w_\a(z) = \Omega_m(z)(\g^m \l)_\a\,.
\eeq
The basic gauge-invariant quantities are the current $J_\l$, the energy momentum tensor $T_\l$ and
the Lorentz current $N_{mn}$ given by
\beq\label{basicInv}
J_\l(z) = w_\a \l^\a\,,\qquad T_\l(z) = w_\a \p \l^\a\,,\qquad N^{mn}(z) = \half(w\g^{mn}\l)\,.
\eeq
Since the conjugate pair $(\l^\a,w_\a)$ is not free due to the pure spinor constraint, the OPEs
of these gauge invariants are computed using the $U(5)$ parameterization of $\l^\a$, with the
$SO(10)$-covariant result \cite{psf},
\begin{align}
N^{mn}(z)\lambda^{\alpha}(w) &\sim {{1\over2}(\gamma^{mn}\lambda)^{\alpha}(w) \over z-w}\,,
&J_{\lambda}(z)\lambda^{\alpha}(w) &\sim {\lambda^{\alpha}(w)\over z-w}\,, \\
N^{mn}(z)J_{\lambda}(w) &\sim \text{regular}\,,
&J_{\lambda}(z)J_{\lambda}(w) &\sim {-4\over(z-w)^{2}}\,,\cr
N_{mn}(z)T_{\lambda}(w) &\sim {N_{mn}(w)\over (z-w)}\,,&J_{\lambda}(z)T_{\lambda}(w) &\sim
{-8\over(z-w)^{3}}+{J_{\lambda}(w)\over (z-w)^{2}}\,,\notag
\end{align}
\begin{align}
N^{mn}(z)N_{pq}(w) &\sim {N^m{}_p\d^n_q-N^n{}_p\d^m_q + N^n{}_q\d^m_p-N^m{}_q\d^n_p \over z - w}
- {3(\d^n_{p}\d^m_{q}-\d^n_q\d^m_p) \over (z - w)^2}\,,\notag\cr
T_{\lambda}(z)T_{\lambda}(w)
&\sim {11\over(z-w)^{4}}+{2T_{\lambda}(w)\over(z-w)^{2}}
  + {\p T_{\lambda}(w)\over z-w} \,.\notag
\end{align}
Similarly, the constraints \eqref{nonminc} imply that the conjugates $\bar w^\a$ and $s^\a$ of conformal weight $+1$
can only appear in gauge-invariant combinations under
\beq
\d\bar w^\a = \bar\Omega_m(\g^m\bar\l)^\a - \phi_m(\g^m r)^\a\,,\qquad
\d s^\a=\phi_m(\g^m\lb)^\a\,,
\eeq
where $\bar\Omega_m$ and $\phi_m$ are arbitrary parameters. The non-minimal counterparts of the
gauge invariants \eqref{basicInv} are given
by \cite{NMPS},
\begin{gather}
\label{nmgauge}
\bar N_{mn}= \half (\bar w\g_{mn}\lb - s\g_{mn} r),\quad
\bar J_{\lb} =\bar w^\a \bar\l_\a -s^\a r_\a,\quad
T_\lb = \bar w^\a\p\bar\l_\a - s^\a \p r_\a,
\end{gather}
with additional gauge invariants
\beq\label{nmgII}
S_{mn} = \half s \g_{mn} \lb,\quad  S=  s^\a \lb_\a\,.
\eeq
The above gauge invariants are related via the BRST charge
\beq
\bar N_{mn} = QS_{mn},\quad \bar J_\lb = QS,\quad T_\lb = Q(s^\a\p\lb_\a)\,,
\eeq
Therefore, the operator $\bar q=\oint \bar J_\lb$ counting the non-minimal variables is BRST
exact, and
satisfies $\bar q \lb_\a = \lb_\a$ and $\bar q r_\a = r_\a$. Therefore if a BRST-closed state $Q\Psi=0$ has
non-vanishing non-minimal $\bar q$ charge $\bar q\Psi=n\Psi$ with $n\neq0$, it is also BRST-exact; $\Psi={\bar
q\over n}\Psi$. And since $S$ and $S^{mn}$ are not closed and $(\bar N_{mn},  \bar J_\lb, T_\lb)$ are
exact, the quartet $(\bar
w^\a,\lb_\a,s^\a,r_\a)$ of non-minimal variables decouples from
the cohomology in what is known as the Kugo-Ojima quartet mechanism.

Moreover,
the energy-momentum tensor,
\beq\label{EMt}
T(z)=-{1\over \ap}\p x^m \p x_m -
p_\a \p\t^\a + w_{\a}\p\l^{\a} +
\wb^\a \p\lb_{\a} -s^\a\p r_{\a} \, ,
\eeq
is related to the BRST charge through the $b$ ghost as $\{Q,b(z)\} = T(z)$, where
\begin{align}
\label{bghost}
b  &= s^\a \p\lb_\a +
{1\over 4(\l\lb)}\bigl[ 2\Pi^{m}(\lb \g_{m}d) -
N_{mn}(\lb \g^{mn}\p\t) - J_{\lambda}(\lb \p\t) - (\lb \p^2 \t) \bigr]\cr
&{} + {(\lb\g^{mnp}r)\over 192 (\l\lb)^2}\Bigl[ {\ap\over 2}(d\g_{mnp}d) + 24N_{mn}\Pi_{p}\Bigr]
\cr
&{}- {\ap\over 2}{(r\g_{mnp}r)(\lb\g^{m}d)N^{np} \over 16(\l\lb)^3 }
+{\ap\over 2} {(r\g_{mnp}r)(\lb\g^{pqr}r)N^{mn}N_{qr} \over 128(\l\lb)^4} \,.
\end{align}
After extracting the non-minimal $U(1)$ ghost-number current
\beq\label{nJ}
J(z) = w_\a \l^\a - s^\a r_\a - {2\big((\lb\p\l)+(r\p\t)\big)\over\llb} +{2(\l r)(\lb\p\t)\over \llb^2}
\eeq
from the double pole of the $b$ ghost with the integrand $\l^\a d_\a + \bar w^\a r_\a$
of the BRST charge,
the
non-minimal pure spinor formalism was shown
in \cite{NMPS} to be a critical $N=2$ topological string.
More precisely, using the terminology $G^+(z)= (\l^\a d_\a + \bar\om^\a r_\a)$, $G^-(z) = b(z)$
one can show that
$T(z)$, $G^+(z)$, $G^-(z)$  and $J(z)$ satisfy the OPEs \cite{NMPS,b2rennan}
\begin{align}
T(z)T(w) &\sim
\frac{2T}{(z-w)^2} + \frac{\p T}{(z-w)}\\
T(z)G^+(w) &\sim \frac{G^+}{(z-w)^2}+ \frac{\p G^+}{(z-w)}\cr
T(z)G^-(w) &\sim \frac{2G^-}{(z-w)^2}+ \frac{\p G^-}{(z-w)}\cr
G^+(z)G^-(w) &\sim \frac{3}{(z-w)^3} +
\frac{J}{(z-w)^2} + \frac{T}{(z-w)}\cr
T(z)J(w) &\sim \frac{-3}{(z-w)^3} + \frac{J}{(z-w)^2} + \frac{\p J}{(z-w)}\cr
J(z)G^\pm(w) &\sim \pm \frac{G^\pm}{(z-w)}\cr
J(z)J(w) &\sim \frac{3}{(z-w)^2}\cr
G^\pm(z)G^\pm(w) &\sim \text{regular},\notag
\end{align}
which identifies them as the generators of a $\hat c=3$ $N=2$ twisted topological conformal algebra.
As such, not only the BRST charge has to be nilpotent but also the $b$
ghost (see e.g. \cite{farrill}). A proof that $b^2=0$ with \eqref{bghost} can be found in \cite{b2rennan,b2chandia}.
Note that the simpler
BRST-equivalent $U(1)$ charge $J(z)=w_\a\l^\a - \bar w^\a\lb_\a$
was show in \cite{NMPS} to
preserve the essential features of the topological string and therefore can be used instead of \eqref{nJ} to
define the ghost-number of pure spinor operators.

\subsubsection{Vertex operators and amplitude prescription}

Vertex operators for massless open-string states are constructed from the linearized
SYM superfields of \eqref{singleS} as
\begin{align}
\label{vertex}
V&=\l^\a A_\a(x,\t),\\
U &= \p \theta ^\a A_\a (x, \theta) + \Pi ^m A_m(x,\theta) + d_\a W^\a(x, \theta) + \half N_{mn}
F^{mn} (x, \theta)\notag
\end{align}
and are independent on the non-minimal variables using the quartet mechanism discussed above.
$V$ is called the unintegrated vertex and has conformal weight zero and $U$ is called the integrated
vertex and has conformal weight ${+}1$. They are related via the BRST charge \eqref{BRST} by
$QU=\p V$, so the integrated vertex is BRST closed up to a total derivative on the worldsheet.
The unintegrated vertex is BRST closed as a consequence of \eqref{dK}, the equation of motion
\eqref{singleS}, as well as the pure spinor constraint \eqref{PSconstraints}
\beq\label{QV}
QV = \l^\a\l^\b D_\a A_\b = \half \l^\a \g^m_{\a\b}\l^\b A_m = 0\,.
\eeq
Their closed-string versions are obtained by a double-copy of the open-string 
vertex operators with the plane-wave factor stripped off, that is $|V|^2 = V(\t){\tilde
V}(\tilde\t)e^{k\cdot x}$ where $V(\t)=\l^\a A_\a(\t)$ with $A_\a(\t)$ as in \eqref{lintheta}, and similarly
for $|U|^2$.

The prescription to calculate $n$-point closed-string amplitudes at genus $g$ is
\begin{align}
\label{presctree}
{\cal A}_{g=0} &= \k^n e^{-2\l}\int_\Sigma \prod_{j=2}^{n-2} d^2z_j |\langle {\cal N}_0
V_1(0)U_j(z_j) V_{n-1}(1)V_n(\infty)\rangle|^2\\
\label{presc1loop}
{\cal A}_{g=1} &= {1\over2}\k^n \int_{\Sigma,{\cal M}_1} d^2\tau_1 \prod_{j=2}^n d^2z_j |\langle {\cal N}
(b,\mu_1) V_1(0) U_j(z_j) \rangle|^2\\
\label{presc2loop}
{\cal A}_{g>1} &= \k^n e^{2\l}(1-\half\d_{g,2})\int_{\Sigma,{\cal M}_g}  d^{3g-3}\tau \prod_{j=1}^n d^2z_j |\langle 
U_j(z_j)\prod_{I=1}^{3g-3}(b,\mu_I) {\cal N}\rangle|^2
\end{align}
where $U(z)$ is the integrated vertex operator \eqref{vertex}, $\tau_I$ for $I=1, \ldots,3g-3$ are
the complex Teichm\"uller parameters with $\mu_I$ their associated Beltrami differentials, the $b$ ghost is given by
\eqref{bghost} and
\beq\label{bmu}
(b,\mu_I) = {1\over2\pi}\int d^2 z b_{zz}\mu_I^z{}_{\bar z}\,,
\eeq
${\cal N}$ is the regularization factor \eqref{Nreg} responsible for convergence as
$(\l\lb)\to\infty$, $\k$ is the normalization
of the vertex operators ($\k^2 = e^{2\l}\pi/\ap^2$ by unitarity) and $e^{2(g-1)\l}$ is the string
coupling constant as in \cite{DHokerS}. The factor of $1/2$ in the genus-two amplitude is required
because all genus-two curves have a $\mathbb{Z}_2$ symmetry \cite{kodaira}.
In addition, $|.|^2$ signifies the holomorphic square of the integrand with the
plane waves of the vertex operators dealt with as described above, and it is important to
emphasize that all calculations are done in the left- and right-moving sectors separately using the
chiral splitting formalism explained below.

\medskip
\noindent\textbf{Integration of non-zero modes}
The OPEs
in a genus $g$ Riemann surface
are used to integrate out the non-zero modes
of the fields of conformal weight $+1$. To do this, we first separate off the zero modes as
(using $d_\a(z)$ to illustrate the procedure)
\beq
d_\a (z) = \hat d_\a (z) + \sum_{I=1}^g d_\a ^I \, \om _I(z)\,,\qquad
\oint _{A_I} \hat d_\a=0
\label{zvsnonz}
\eeq
where $\om_I(z)$ are $g$ holomorphic one-forms satisfying $\oint_{A_I}\om_J(z)dz = \d_{IJ}$ and
$A_I$ represents the $A$ cycles of the Riemann surface. Then the non-zero modes (indicated by hats) are integrated out
via their OPEs.
For example,
\begin{align}
\hat p_\alpha(z) \, \theta^\beta(y) &\sim \partial_z \ln E(z,y) \, \delta_\alpha^{\beta} 
\notag \\
\hat d_\alpha(z) \, K \big(x(y),\t(y) \big ) &\sim  \p_z \ln E(z,y) \, D_\alpha K \big  (x(y), \theta (y) \big )
\label{OPEwithE} \\
\hat \Pi_m(z) \, K\big(x(y),\t(y) \big ) &\sim -\p_z \ln E(z,y)  \, \partial_m K \big (x(y), \theta (y)\big )
\notag
\end{align}
where $E(z,y)$ is the prime form and
$K(x,\theta)$ is an arbitrary superfield depending on $x$ and $ \theta$, but not on the worldsheet derivatives of these fields.
In the limit where $z\to y$, the prime form behaves as $E(z,y)\sim z-y$ and the propagator $\p_z
\ln E(z,y)$ displays its distinctive singular structure $\sim 1/(z-y)$ seen in \eqref{OPEs}.
The OPE of the $x^m(z,\bar z)$ fields
\beq
X^m(z,\bar z)X_n(w,\bar w) \sim -{\ap\over2}\d^m_n G(z,w),
\eeq
with $G(z,w)$ the genus-$g$ Green function,
couples the left- and right-movers and motivates the chiral
splitting techniques developed by D'Hoker and Phong.

\medskip
\noindent\textbf{Zero-mode integrations}
The zero-mode integrations that remain after integrating out the non-zero
modes via OPEs are performed using
\beq\label{zerom}
\langle \ldots \rangle = \int [d\theta][dr][d\l][d\lb] \prod_{I=1}^g [dd^I][ds^I][d\wb^I][dw^I] \ldots
\eeq
where \cite{2loopH}
\begin{align}
&[d\l]\, T_{\a_1 \ldots\a_5} = c_{\l}\, (\e\cdot d^{11}\l)_{\a_1 \ldots\a_5}\,,
&[dw] &= c_{w}\,(T\cdot\e\cdot d^{11}w)\cr
&[d\lb]\, {\bar T}^{\a_1 \ldots\a_5} \kern-2pt = c_{\lb}\,(\e\cdot d^{11}\lb)^{\a_1 \ldots\a_5}\,,
&[dr] &= c_r\, ({\bar T}\cdot\e\cdot\p^{11}_r)\cr
&[d\wb]\, T_{\a_1 \ldots\a_5} = c_{\wb}\, (\e\cdot d^{11}\wb)_{\a_1 \ldots\a_5}
&[ds^I] &= c_s\, (T\cdot\e\cdot \p^{11}_{s^I})\cr
&[d\t] = c_\t\, d^{16}\t
&[dd^I] &= c_d\, d^{16}d^I.
\end{align}
with the shorthand $(\e\cdot d^{11}\l)_{\a_1 \ldots\a_5}:={1\over11!}\e_{\a_1 \ldots\a_{16}} d\l^{\a_6}\kern-4pt \ldots
d\l^{\a_{16}}$, and its contraction
$({\bar T}\cdot\e\cdot d^{11}\l)={1\over11!5!}{\bar T}^{\a_1 \ldots\a_5}(\e\cdot d^{11}\l)_{\a_1 \ldots\a_5}$
with similar expressions for the others.
The expressions of $T_{\a_1\ldots \a_5}$ and ${\bar T}^{\a_1\ldots \a_5}$ are given by
\begin{align}
T_{\a_1\a_2\a_3\a_4\a_5} &=
(\l \g^m)_{\a_1}(\l \g^n)_{\a_2}(\l \g^p)_{\a_3} (\g_{mnp})_{\a_4\a_5} \\
{\bar T}^{\a_1\a_2\a_3\a_4\a_5} &= (\lb \g^m)^{\a_1}(\lb \g^n)^{\a_2}(\lb \g^p)^{\a_3}
(\g_{mnp})^{\a_4\a_5}\notag
\end{align}
and they
can be shown to be totally antisymmetric due to the pure spinor constraints and satisfy
$T\cdot {\bar T} = 5!\, 2^6 (\l\lb)^3$.
Finally, the normalizations are given by
\begin{align}
 c_{\l} &= \halfap{-2}\mkern-4mu \Big({A_g \over 4\pi^2}\Big)^{\!\! 11/2}
 &c_{w} &= \halfap{2} {1\over (2\pi)^{11} Z_g^{11/g}}\\
c_{\lb} &= 2^6\halfap{2}\Big({A_g \over 4\pi^2}\Big)^{\mkern-6mu 11/2}
&c_r &= R\halfap{-2}\Big({2\pi \over A_g}\Big)^{\mkern-6mu 11/2}\cr
c_{\wb} &= \halfap{-2} {(\l\lb)^3\over (2\pi)^{11}} Z_g^{-11/g}
&c_s &= \halfap{2} {(2\pi)^{11/2} \over 2^6 R(\l\lb)^3} Z_g^{11/g}\cr
 c_{\t} &= \halfap{4}\Big({2\pi\over A_g}\Big)^{\mkern-6mu 16/2}
&c_{d} &= \halfap{-4} \!\! (2\pi)^{16/2}\, Z_g^{16/g}\,, \notag
\end{align}
where $A_g = \int d^2z \sqrt{g}$ is the area of the genus-$g$ Riemann surface. Moreover,
$Z_g = 1/\sqrt{\det(2\ImOmega)}$ where $\Omega_{IJ}$ is the period matrix,
and $R$ is a free parameter that is used to choose the normalization of the three-point amplitude
at genus zero (after which the normalization of all genus-$g$ $n$-point
amplitudes is fixed).
As shown in \cite{2loopH}, the closed-string amplitudes are independent on the
area $A_g$ because the number of bosonic and fermionic variables of conformal weight $0$ is the
same, and independent on the choice of normalization of the holomorphic one-forms
because the number of bosonic and fermionic variables of conformal weight $+1$ is the same.
The factor
\beq\label{Nreg}
{\cal N} =  e^{-(\l\lb) - (r\t) + \sum_{I=1}^g (s^Id^I) - (w^I\wb^I)  }.
\eeq
regulates the zero-mode integrations over the non-compact spaces of
the bosonic variables $\l^\a, \lb_\a$ and $w^\a,\bar w_\a$ as $\llb\to\infty$ and $(w\bar
w)\to\infty$ in a
manner explained in \cite{NMPS}. The formula for the integration over the pure spinor
variables was found in \cite{humberto} using techniques from algebraic geometry
\beq
\int [d\l][d\lb] (\l\lb)^n e^{-(\l\lb)} = \Big({A_g\over 2\pi}\Big)^{\mkern-6mu
11}{\Gamma(8+n)\over 7!\, 60}\,,
\eeq
where $\Gamma(x)$ is the gamma function. The $b$ ghost \eqref{bghost} has factors of $1/(\l\lb)$
which are not regularized by the regulator \eqref{Nreg} as $\llb\to0$. It was shown in \cite{NMPS}
that as long as the integrands diverge slower than $1/(\l^{8+3g}\lb^{11})$ the amplitudes are still
well-defined due to a compensating factor of $\l^{8+3g}\lb^{11}$ arising from $\langle {\cal N}
\ldots\rangle$ in \eqref{zerom}. As explained in \cite{NMPS}, this issue is closely related to the existence of the operator
\beq\label{xiop}
\xi = {(\l\t)\over \llb + (r\t)} = {(\l\t)\over \llb}\sum_{n=0}^{11}\left({(r\t)\over \llb}\right)^n
\eeq
where the Taylor expansion ends at $n=11$ because there are only 11 degrees of freedom in $r_\a$
due to the constraint \eqref{nonminc}. This operator trivializes the cohomology as $Q\xi=1$ but
$\langle {\cal N}\xi(\l^3\t^5)\rangle$ diverges faster than $1/(\l^{8+3g}\lb^{11})$, therefore if
the integrands were allowed to diverge too fast they would also be BRST-exact. Forbidding such
pathological behavior restricts the amplitude prescription to contain at most three $b$ ghosts, or
in other words, up to genus two.
By regularizing the $b$ ghost to remove the singularity as $\llb\to0$, an alternative prescription
that allows amplitudes at arbitrary genus to be well defined was proposed in \cite{BerkovitsNekrasov}.

As emphasized in \cite{2loopH}, after the integration over $[dd^I][ds^I][dw^I][d\wb^I]$ has been performed, the remaining
integrations over
$\l^\a,\lb_\b,\t^\d$ and $r_\a$ are the same ones appearing
in the prescription of the tree-level amplitudes, and therefore give rise to (non-minimal) pure spinor superspace
expressions.

\medskip
\noindent\textbf{Chiral splitting}
To address the mixing of left- and right-movers via OPE contractions -- an issue that prevents
writing the closed-string correlator as an holomorphic square -- the chiral splitting procedure
\cite{DHoker:1988pdl,DHoker:1989cxq,Verlinde:1987sd} introduces loop momenta $\ell^m_I$
\beq\label{loopmom}
\ell^m_I = \oint_{A_I} dz \Pi^m(z)
\eeq
in order to rewrite conformal correlators of the $x^m$-field in terms of an integral over $\ell_I$.
The
integrand then becomes a product of left- and right-movers of schematic form $\cF_n(z_i,k_i,\ell^I)
\overline{ \tilde \cF_n (z_i, - \bar k_i, -\ell^I)}$, denoted {\it chiral blocks}.
Chiral blocks have a universal monodromy behavior as the points are moved
around one another or circled around the homology cycles of the surface, and these properties
can be exploited\footnote{Of course, the monodromy of the chiral blocks play a central role in
calculations with the RNS formalism, see e.g.\cite{2loopDP}, but in this review we will focus on the pure spinor formalism.}
to propose pure spinor superstring integrands \cite{oneloopIII,2loopI}. More precisely,
decomposing the chiral blocks into chiral kinematic correlators $\cK(z_i,\ell^I)$ and a chiral
Koba-Nielsen factor $\cI_n$ (to be displayed below) as $\cF = \langle\cK_n\rangle \cI_n$, the
expression for the chiral correlator must be
invariant under the combined {\it homology shifts} of vertex positions $z_i$ and loop momenta
$\ell^I$ around the $A_I$ or $B_I$ cycles:
\begin{align}
\label{homshift}
\cK_{n}(z_i,k_i,\ell^I)  & =   \cK_n(z_i + \delta_{ij} A_J , k_i,\ell^I)\\
 \cK_n(z_i,k_i,\ell^I)  & =
 \cK_n(z_i+ \delta_{ij} B_J ,  k_i, \ell^I -  2\pi \delta^I_J\,  k_j )\,.\notag
\end{align}
When viewed as a constraint on the chiral correlator, these invariances can be
used as a guide to obtain superstring correlators
\cite{oneloopI,oneloopII,oneloopIII,2loopI,2loopII}.

\subsubsection{Pure spinor superspace}

After all the non-zero modes of the worldsheet fields have been integrated out using OPEs, the
correlator contains only the zero modes of conformal-weight zero variables. In the minimal
pure spinor formalism of \cite{MPS} that means the zero modes of $\l^\a$ and $\t^\a$, while in the non-minimal
formalism they can also include $\lb_\a$ and $r_\a$ variables. In the latter case, one can show that
$r_\a$ can be converted to supersymmetric derivatives $D_\a$ while the pure spinors $\lb_\a$ can always
be arranged to contract $\l^\a$ to produce scalar factors of $(\l\lb)$ which change the
normalization factor. Therefore the zero mode
integration (with a constant number of $(\l\lb)$ factors) can be done with the prescription
\cite{psf}
\beq\label{PSS}
\langle (\l\g^m\t)(\l\g^n\t)(\l\g^p\t)(\t\g_{mnp}\t)\rangle = 2880\,.
\eeq
This motivates the notion of {\it pure spinor superspace} \cite{PSS}, defined as expressions containing three
pure spinors and an arbitrary number of SYM superfields composed of polarizations, momenta and
$\t^\a$ variables. The prescription \eqref{PSS} justifies the previous claim that terms of order
$\t^{>5}$ in \eqref{lintheta} could be safely ignored. As a simple example of pure spinor
superspace one can consider extracting the supersymmetric expression of the
massless open-string three-point amplitude at genus zero,
\begin{align}\label{VVV}
\langle V_1 V_2 V_3 \rangle &=
\Big[{1\over 64}
k^2_m e^1_{r}e^2_{n} e^3_{s}
\langle (\l \g^r\t) (\l \g^s\t)(\l \g_p \t)(\t\gamma^{pmn}\t)\rangle\cr
&\qquad{} + {1\over18}e^m_1 \langle(\l\g_m\t)(\l\g_n\t)(\l\g_p\t)(\t\g^n\chi_2)(\t\g^p\chi_3)\rangle
+ {\rm cyclic}(1,2,3)\Big]\cr
&=\half e^m_1 f^{mn}_2 e^n_3 + e^1_m(\chi_2\g^m \chi_3) + {\rm cyclic}(1,2,3)\,.
\end{align}
where we plugged in the $\t$ expansions of \eqref{lintheta} and kept only the terms with $\t^5$.
Moreover, we used
\begin{align}
\langle (\l \g_r\t) (\l \g_s\t)(\l \g_p \t)(\t\gamma^{pmn}\t)\rangle
&= 64\d_{rs}^{mn},\\
\langle(\l\g^m\t)(\l\g_n\t)(\l\g_p\t)(\t\g^n\chi_2)(\t\g^p\chi_3)\rangle&=
18(\chi_2\g^m\chi_3),\notag
\end{align}
which can be derived from group-theory considerations (see appendix of \cite{anomaly}),
momentum conservation $k^m_1+k^m_2+k^m_3 = 0$ and the transversality condition $(k_i\cdot e_i) = 0$.

\subsubsection{Multiparticle superfields}

While four-point scattering amplitudes at one and two loops can be
written down using the (single-particle) SYM superfields,
the OPE contractions present at higher points lead to linear combinations
of SYM superfields whose patterns are captured by so-called {\it multiparticle superfields},
describing multiple string states at the same time.
Not only they encode the numerators associated to OPE singularities, but they are also designed
in a way that removes BRST-exact pieces and total derivatives. The end result
displays covariant BRST
transformations and
generalized
Jacobi identities \cite{genjac} -- the latter property is particularly useful for
describing Bern-Carrasco-Johansson
color/kinematics duality \cite{BCJ}.

The two-particle superfields generalizing the standard superfields of \eqref{singleS}
are given by \cite{EOMbbs}
\begin{align}
A^{12}_\a &=  \tfrac{1}{2}\bigl[ A^2_\a (k_2\cdot A_1) + A_2^m (\g_m W_1)_\a - (1\leftrightarrow 2)\bigr] \, ,\cr
A_{12}^m &=  \tfrac{1}{2}\bigl[  A_2^m(k_2\cdot A_1) + A^1_p F_2^{pm} + (W_1\g^m W_2) - (1\leftrightarrow
2)\bigr] \, ,\cr
W_{12}^\a &= \tfrac{1}{4}(\g_{mn}W_2)^\a F_1^{mn} + W_2^\a (k_2\cdot A_1) - (1\leftrightarrow 2) \, ,\label{twopart} \\
F_{12}^{mn} &= F_2^{mn}(k_2 \cdot A_1)+ \tfrac{1}{2} F_{2}^{[m}{}_p F_{1}^{n]p}
+ k_1^{[m}(W_1 \gamma^{n]} W_2) - (1\leftrightarrow 2) \,,\notag
\end{align}
and satisfy
\begin{align}
\label{twoEOM}
D_{\a} A^{12}_{\b} + D_{\b} A^{12}_{\a} &= \g^m_{\a\b}A^{12}_m + (k_1\cdot k_2)(A^1_\a A^2_\b + A^1_\b A^2_\a) \, , \\
D_\a A_{12}^m &= \g^m_{\alpha \beta} W_{12}^{\beta} \!+\! k_{12}^m A^{12}_\a \!+\! (k_1\cdot k_2)(A^1_\a A_2^m \!-\! A^2_\a A_1^m) \, ,\cr
D_\a W^\b_{12}&= \tfrac{1}{4}(\g_{mn})_\a{}^\b F_{12}^{mn} + (k_1\cdot k_2)(A^1_\a W_2^\b - A^2_\a W^\b_1)\, ,\cr
D_\a F_{12}^{mn}&= k_{12}^{[m} (\g^{n]} W_{12})_\a  + (k_1\cdot k_2)\big[A^1_\a F_2^{mn}  +  A_{1}^{[n} (\g^{m]} W_2)_\a - (1\leftrightarrow 2)\big]\, .\notag
\end{align}
These equations of motion have the same form as in the single-particle case
\eqref{singleS} with additional corrections proportional to $(k^1\cdot k^2)$.
The construction of (local) multiparticle superfields of arbitrary multiplicity leads to superfields
labelled by words $P=p_1p_2p_3 \ldots$ or by arbitrary nested commutators $P=[ \ldots[[p_1,p_2],p_3],
\ldots]$ (e.g. $A^m_{1234}$ or $F^{mn}_{[1,[2,3]]}$) and can be found in
\cite{EOMbbs,genredef}.

\subsection{Superstring amplitudes with pure spinors}

The pure spinor prescription to compute genus-$g$ amplitudes relies
on the basic fact that the OPE analysis of primary operators
determines a meromorphic function of the vertex positions due to its poles and residues.
In the absence of monodromy such as at genus zero,
this completely determines the correlator, but this is no longer true at higher genus.
On a surface of higher genus, the existence of holomorphic one-forms
implies that the knowledge of the positions and residues of the
poles from the OPE analysis no longer suffices to completely determine the correlator; the regular terms
contain non-trivial information.
In principle, the zero modes
provide the additional information to find the complete correlator \cite{chiralBosonization}.
However, sometimes this is impractical to follow systematically and the calculation benefits
from the practical
requirements of {\it homology} and {\it BRST} invariance\footnote{It is worth mentioning
that several amplitudes computed in this manner used the ``minimal'' pure
spinor formalism and its simpler pure spinor superspace expressions depending only on the zero
modes of $\l^\a$ and
$\t^\a$ (the expressions in the non-minimal formalism also depend on $\lb_\a,
r_\a$ in intermediate stages).}
constraints to be discussed below.

\bigskip
\noindent\textbf{Homology invariance} The introduction of loop momentum integrals with the chiral splitting formalism
had to pass the consistency check that the integrated amplitudes were single-valued as a function
$\hat f(z_i)$
of the vertex positions $z_i$ {\it after} the
loop momentum was integrated out \cite{DHoker:1989cxq}.
However, a stronger constraint was proposed\footnote{It was initially dubbed
``monodromy invariance'' and it led to the development of {\it
generalized elliptic integrands} (GEI) in the context of genus-one string
amplitudes \cite{oneloopII}.}
in \cite{1loopDouble,oneloopII}: that the chiral
{\it integrands}, viewed as a function $f(\ell_I,z_i)$ of both the loop momenta $\ell_I$ and vertex
positions $z_i$ should be strictly single-valued
under the monodromies of the loop momentum and the vertex positions as they move around $A_I$ and
$B_I$ cycles:
\beq\label{GEI}
f(z'_i,\ell'_I) = f(z_i,\ell_I),\qquad \begin{cases}
\text{$A_I$-cycle}:& (z'_i,\ell'_I) = (z_i+\delta_{ij}A_J,\ell_I)\cr
\text{$B_I$-cycle}:& (z'_i,\ell'_I) = (z_i+\delta_{ij}B_J,\ell_I-2\pi i\delta^J_I k_j)
\end{cases}
\eeq
That is, the chiral integrands should be single-valued {\it before} the loop momentum is integrated out. This requirement
interlocks the different sectors of the integrands with different powers of loop momenta
with predictive consequences:
it can be used to constrain and obtain the superstring integrands themselves.

This requirement of {\it homology invariance} was used in
\cite{oneloopI,oneloopII,oneloopIII} to determine the integrands of the five, six and seven-point
massless amplitudes at genus one, and in \cite{2loopI} to obtain the massless five-point
integrand at genus two.

\bigskip
\noindent\textbf{BRST invariance}
Superstring scattering amplitudes must be spacetime supersymmetric and gauge invariant. As
explained in detail in \cite{psf}, the cohomology prescription \eqref{PSS} to integrate out the pure spinor
zero modes leads to gauge-invariant and supersymmetric expressions if the pure spinor superspace
expression is BRST invariant.
Recall that when the BRST charge \eqref{BRST} acts on superfield expressions containing only
$x^m$, $\t^\a$ (and
possibly $\l^\a$), the OPE \eqref{OPEwithE} implies
\beq\label{QK}
QK(x,\t) = \l^\a D_\a K(x,\t)
\eeq
where $D_\a$ is the superspace derivative \eqref{covD} and $\l^\a$ is the pure spinor. As we will
see, this
equation plays an important role in the study of the BRST cohomology properties of string scattering
amplitudes.
More precisely, if the outcome of the OPEs
among the vertices is written in pure spinor superspace as $\langle\l^\a\l^\b\l^\g f_{\a\b\g}(
e_i,k_i,\xi_i, \t)\rangle$ where $e_i,\xi_i$ and $k_i$ represent a collection of bosonic and
fermionic polarizations and their momenta, then the amplitude prescription will give rise to gauge
invariant and supersymmetric expressions if
\begin{align}
Q\big(\l^\a\l^\b\l^\g f_{\a\b\g}(e_i,k_i,\xi_i, \t)\big) &= 0\,,\\
\l^\a\l^\b\l^\g f_{\a\b\g}(e_i,k_i,\xi_i, \t)&\neq Q\Omega.\notag
\end{align}
This implies that the superspace expressions of arbitrary scattering amplitudes must be in the
{\it cohomology of the BRST charge}. This requirement together with the OPE structure of the
genus-zero pure spinor prescription  is enough to completely
determine the tree-level scattering amplitudes of ten-dimensional SYM theory \cite{towardsFT,nptMethod}.

\subsubsection{Genus zero}

\medskip
\noindent\textbf{SYM tree amplitudes}
The knowledge that the genus-zero superstring amplitudes reduce to ten-dimensional SYM tree
amplitudes \cite{neveuscherkFT} has a powerful consequence: the tree amplitudes in field theory have the same superfield structure
as their string theory counterparts. This led to the suggestion that the BRST cohomology structure
of pure spinor superspace 
expressions inspired by the pure spinor prescription could be used to completely fix the form of
the SYM tree amplitudes \cite{towardsFT}. Using multiparticle superfields, the first non-vanishing
tree amplitudes were found to be
\begin{align}
\label{3ptSYM}
A(1,2,3) &= \langle V_1 V_2 V_3\rangle\,, \\
A(1,2,3,4) &=
{\langle V_{12}V_3V_4\rangle\over s_{12}}
+ {\langle V_{1}V_{23}V_4\rangle\over s_{23}}\,, \notag\\
A(1,2,3,4,5) &=
  {\langle V_{123}V_{4}V_{5}\rangle\over s_{12} s_{45}}
+ {\langle V_{321}V_{4}V_{5}\rangle\over s_{23} s_{45}}
+ {\langle V_{12} V_{34}V_{5}\rangle\over s_{12} s_{34}}
+ {\langle V_1 V_{234}V_{5}\rangle\over s_{23} s_{51}}
+ {\langle V_1 V_{432}V_{5}\rangle\over s_{34}s_{51}} \,.\notag
\end{align}
The regular structure of the BRST variation of certain non-local multiparticle superfield
building blocks $M_P$, the Berends-Giele currents, and various other hints led to the general
$n$-point expression for SYM tree amplitudes in \cite{nptMethod}:
\beq\label{SYMtree}
A(P,n) = \sum_{XY=P}\langle M_XM_Y M_n\rangle
\eeq
where $XY=P$ represents the sum over all deconcatenations of the word $P$ into the words $X$ and
$Y$ (including the empty word provided we define $M_\emptyset :=0$). In this language, the
amplitudes \eqref{3ptSYM} become
\begin{align}
\label{BG3ptSYM}
A(1,2,3) &= \langle M_1 M_2 M_3\rangle \, ,\\
A(1,2,3,4) &= \langle M_{12}M_3M_4\rangle + \langle M_{1}M_{23}M_4\rangle\, ,\notag\\
A(1,2,3,4,5) &=
  \langle M_{123}M_{4}M_{5}\rangle
+ \langle M_{12} M_{34}M_{5}\rangle
+ \langle M_1 M_{234}M_{5}\rangle\,.\notag
\end{align}
The explicit expressions for the Berends-Giele currents in terms of multiparticle superfields, the
first of which are given by
\beq
M_1 = V_1\,,\qquad
M_{12} = {V_{12}\over s_{12}}\,,\qquad
M_{123} = {V_{123}\over s_{12}s_{123}} + {V_{321}\over s_{23}s_{123}}\,,
\eeq
can be constructed in a multitude of ways (see \cite{pstreereview}). Their
BRST variation admits
a simple all-order form
\beq\label{QMP}
QM_P = \sum_{XY=P}M_XM_Y\,,
\eeq
from which it easily follows that the superfield expression in \eqref{SYMtree} is BRST closed. It
is also
not BRST exact, and therefore it is in the cohomology of the BRST charge. To see this, note that $M_P$ contains a divergent propagator $1/s_P$ 
in the phase space of $|P|{+}1=n$ massless particles, so one cannot write the superfields in
\eqref{SYMtree} as $Q(M_P M_n)$. In other words, $M_PM_n$ is not an allowable BRST ancestor, which
explains why $\langle \sum_{XY=P}M_XM_YM_n\rangle\neq0$.

\medskip
\noindent\textbf{The $n$-point superstring disk correlator}
The general $n$-point disk correlator of massless string states was computed in \cite{nptStringI} using
multiparticle superfield techniques to capture the OPE singularities of vertex operators. The result
can be written as a sum over $(n-3)!$ SYM field-theory tree amplitudes \eqref{SYMtree} as follows
\beq\label{nptdisk}
{\cal A}_n(P)= (2\ap)^{n{-}3}\int d\mu_P^n
\bigg[ \prod_{k=2}^{n{-}2}\sum_{m=1}^{k-1} {s_{mk} \over z_{mk}}\; A(1,2,\ldots,n)  +
\perm(2,3,\dots,n{-}2) \bigg]\,,
\eeq
where $\int d\mu_P^n$ is a shorthand for the integration over the vertex positions with integration
domain $D(P)$ and weighted by
the genus-zero Koba-Nielsen factor
$\int_{D(P)} \prod_{j=2}^{n{-}2} dz_j \prod_{1\leq i<j}^{n-1} |z_{ij}|^{- 2\alpha' s_{ij}}$. This
result motivated the development of a method \cite{drinfeld} to obtain the $\ap$ expansion of the
integrals in \eqref{nptdisk}. In addition, the conclusion that there is a $(n-3)!$ basis of tree amplitudes in the
work of Bern, Carrasco and Johansson \cite{BCJ} becomes manifest as the left-hand side must reduce, in the
limit as $\ap\to0$, to
a color-ordered SYM tree amplitude $A(P)$ with arbitrary ordering $P$, which in turn is expanded
in terms of  $(n-3)!$ tree amplitudes on the right-hand side.
For an in-depth discussion of these matters, see \cite{pstreereview}.

\subsubsection{Genus one}

We are now going to showcase some of the results obtained with the pure spinor formalism
at genus one. For the open string, the amplitudes have the general form
\beq\label{openagain}
{\cal A}_n  =
\sum_{\rm top} C_{\rm top} \int_{D_{\rm top}}\!\!\!\!
d\tau \, dz_2 \, d z_3 \,\ldots\, d z_{n} \, \int d^{D} \ell \ |{\cal I}_n(\ell)|\,
\langle {\cal K}_n(\ell)\rangle\,,
\eeq
with $\langle\ldots \rangle$ denoting the zero-mode integration prescription \eqref{zerom},
which will be presented in the examples below as pure spinor superspace expressions in terms
of zero modes of $\l^\a$ and $\t^\a$.
The integration domains $D_{\rm top}$ for the
modular parameter $\tau$ and vertex positions $z_j$ must be chosen according to the
topologies of a cylinder or a M\"obius strip with associated color factors
$C_{\rm top}$. The integration over loop momenta
$\ell$ must be performed as a consequence of the chiral-splitting method,
which, in turn, allows to derive massless
closed-string one-loop amplitudes from an integrand of double-copy form
\beq\label{againclosed}
{\cal M}_n  =
 \int_{{\cal F}}
d^2\tau \, d^2z_2 \, d^2 z_3 \, \ldots \, d^2 z_{n} \,
\int d^{D} \ell \ |{\cal I}_n(\ell)|^2 \,
\langle {\cal K}_n(\ell)\rangle \, \langle\tilde{\cal K}_n(-\ell)\rangle\,,
\eeq
with ${\cal F}$ denoting the fundamental domain for inequivalent tori with
respect to the modular group.
Both expressions \eqref{openagain} and \eqref{againclosed} involve the universal one-loop Koba--Nielsen factor
\beq\label{IIIKNfactor}
{\cal I}_n(\ell) \equiv  \exp\Big( \sum^n_{i<j} s_{ij} \log \theta_1(z_{ij},\tau)
+  \sum_{j=1}^n  z_j(\ell\cdot k_j) + {\tau \over 4\pi i} \ell^2 \Big)\,,
\eeq
with light-like external momenta $k_j$ and
$s_{ij}\equiv k_i \cdot k_j$ as well as $z_{ij} \equiv z_i - z_j$.

\medskip
\noindent\textbf{The Eisenstein-Kronecker series}
As pointed out above, knowing the singularity structure of the superstring correlators
is not enough to reconstruct the full meromorphic integrand as a function of the vertex positions,
as crucial information from the non-singular parts is needed. In \cite{1loopKE}, a generating
series of worldsheet functions was proposed that contained an infinite tower of functions
$g^{(n)}(z)$ for $n\ge0$
on a complex elliptic curve describing a genus-one surface with modulus $\tau$. These functions turn out to have
the correct properties to capture both the
singular part of superstring correlators with $g^{(1)}$ as well as the non-singular pieces with
$g^{(n)}, n\ge2$. More precisely, these functions are constructed via the Laurent series of the
Eisenstein-Kronecker series $F(z,\a,\tau)$ \cite{brownlevin}
\beq\label{KE}
F(z,\alpha,\tau) \equiv
\frac{\theta_1'(0,\tau)\theta_1(z+\alpha,\tau)}{\theta_1(z,\tau)\theta_1(\alpha,\tau)} =
\sum_{n=0}^\infty \a^{n{-}1}g^{(n)}(z,\tau)
\eeq
where $\t_1(z,\tau)$ is the odd Jacobi theta function
($q=e^{2\pi i\tau}$)
\begin{equation}
\theta_1(z,\tau)\equiv 2iq^{1/8}\sin(\pi z)\prod_{j=1}^{\infty}(1-q^j)\prod_{j=1}^{\infty}(1-e^{2\pi i z}q^j)\prod_{j=1}^{\infty}(1-e^{-2\pi i z}q^j)\ ,
\label{jact}
\end{equation}
satisfying  $\theta_1(z+1,\tau)=-\theta_1(z,\tau)$ and $\theta_1(z+\tau,\tau)=-e^{-\pi
i\tau}e^{-2\pi iz}\theta_1(z,\tau)$ as $z$ is moved around the $A$ or $B$ cycle. 
In addition, $\t_1'(z,\tau)=\p_z\t_1(z,\tau)$.
The functions
$g^{(n)}$ for the first few cases are $g^{(0)}(z,\tau)=1$,
\beq
g^{(1)}(z,\tau)= \partial \log \theta_1(z,\tau) \,,\quad
g^{(2)}(z,\tau)= {1\over 2} \Big[
(\partial \log \theta_1(z,\tau))^2 - \wp(z,\tau) \Big]\,,
\eeq
where
$\wp(z,\tau) = -\p^2\log\t_1(z,\tau) - {\rm G}_2(\tau)$
is the Weierstrass function and ${\rm G}_{2k}(\tau)$ are holomorphic Eisenstein series.

The
function $g^{(1)}(z,\tau)$ is singular as $z\to0$ while all $g^{(n)}(z,\tau)$ with $n\ge2$ are
non-singular in this limit. In addition, all $g^{(n)}(z,\tau)$ are single-valued around the $A$-cycle as
$z\to z+1$ but have non-trivial monodromy around the $B$-cycle as $z\to z+\tau$
\beq\label{monodg}
g^{(n)}(z+\tau,\tau) = \sum_{k=0}^n {(-2\pi i )^k \over k!} g^{(n-k)}(z,\tau)\,.
\eeq
For instance,
$g^{(1)}(z+\tau,\tau) = - 2\pi i$ and
$g^{(2)}(z+\tau,\tau) = - 2\pi i g^{(1)}(z,\tau)  + {1\over 2}(2\pi i)^2$.
The singularity structure of these functions as well as their monodromies
in a genus-one surface provided valuable information to constrain and obtain
\cite{oneloopI,oneloopII,oneloopIII} the
genus-one $n$-point superstring correlators for $n\le7$ using the homology invariance principle
discussed above.
The shorthand $g^{(n)}_{ij}:=g^{(n)}(z_i-z_j,\tau)$ will be used below and it will be convenient
to define a linearized $B$-cycle monodromy operator $D$
\beq\label{Dop}
D = -{1\over 2\pi i} \sum_{j=1}^n \Omega_j \delta_j
\eeq
where $\Omega_j$ are formal variables that capture the $B$-cycle monodromies around $z_j$
generated by the formal operator $\d_j$ with action $\delta_j \ell = - 2 \pi i k_j$ and
$\delta_j g^{(n)}_{jm} = - 2 \pi i g^{(n-1)}_{jm}$ for $n\ge 1$ as well as
$\delta_{j}g^{(0)}_{jm}=0$ and $\delta_{j}g^{(n)}_{im}=0$ for all $i,m\neq j$. As discussed in
\cite{oneloopII}, there is a remarkable duality relating the operator $D$ with the BRST charge $Q$.

\medskip
\noindent\textbf{BRST building blocks}
The other ingredient used to obtain the genus-one superstring correlators was the BRST invariance
property of the integrands. This was addressed by the construction of BRST building blocks with
covariant BRST transformations, using multiparticle superfields techniques in combination with
pure spinor zero-mode analysis and group theory to constrain the appearance of superfields.
This led to the definition of multiple BRST building blocks with different BRST transformation
properties allowing for the construction of BRST invariants in the pure spinor cohomology.

For instance, the zero-mode sector with four $d_\a$ zero modes from the $b$ ghost suggests the
scalar building blocks
\beq\label{TABCdef}
T_{A,B,C} = {1\over 3}(\l\g_m W_A)(\l\g_n W_B)F^{mn}_C +
{\rm cyclic}(A,B,C)\,.
\eeq
in terms of multiparticle superfields labelled by words $A,B,C$. Their
BRST variation following \eqref{QK} is given by ($k_\emptyset\equiv 0$)
\beq\label{QTABC}
QT_{A,B,C} =
\!\!\!\sum_{A=XjY\atop Y=R\shuffle S}\!\!
(k_{X} \cdot k_j)\big[
V_{XR}T_{jS,B,C} - V_{jR}T_{XS,B,C}\big] + (A\leftrightarrow B,C)\,,
\eeq
where $\shuffle$ denotes the shuffle product 
defined iteratively by \cite{reutenauer}
\beq\label{shuffledef}
\emptyset\shuffle P = P\shuffle\emptyset := P\, ,\qquad
iP\shuffle jQ := i(P \shuffle jQ) + j(Q \shuffle iP)\,,
\eeq
for letters $i$ and $j$, words $P$ and $Q$ with
$\emptyset$ representing the empty word. For example, $1\shuffle 23=123+213+231$.

For an illustration of \eqref{QTABC}, the BRST variations of all $T_{A,B,C}$ up to multiplicity five
are given by
\begin{align}
\label{QTsone}
QT_{1,2,3}&= 0\,, \\
QT_{12,3,4} &= (k_1\cdot k_2)\big[V_1T_{2,3,4}-V_2T_{1,3,4}\big]\,,\cr
QT_{123,4,5} &= (k_1\cdot k_2)\big[V_1T_{23,4,5}+ V_{13}T_{2,4,5}
 - V_{2}T_{13,4,5} -V_{23}T_{1,4,5}\big]\cr
&\quad{}+ (k_{12}\cdot k_3)\big[V_{12}T_{3,4,5}-V_{3}T_{12,4,5}\big]\,,\cr
QT_{12,34,5} &= (k_1\cdot
k_2)\big[V_1T_{2,34,5}-V_2T_{1,34,5}\big]+(12\leftrightarrow34)\,.\notag
\end{align}
Other zero-mode contributions from the $b$ ghost give rise to tensorial building blocks
with an arbitrary number of vector indices. For simplicity, the vector building block
has the form
\beq\label{TmABCDdef}
T^m_{A,B,C,D}\equiv \big[A^m_A T_{B,C,D} + (A\leftrightarrow
B,C,D)\big] + W^m_{A,B,C,D}
\eeq
with 
\beq
W^m_{A,B,C,D}= {1\over 12}(\l\g_n W_A)(\l\g_p W_B)(W_C\g^{mnp} W_D) + (A,B|A,B,C,D)
\eeq
with the notation $(A_1,{\ldots } , A_p \,|\, A_1,{\ldots} ,A_n)$ instructing
to sum over all possible ways to choose $p$
elements $A_1,A_2,\ldots ,A_p$ out of the set $\{A_1,{\ldots} ,A_n\}$, for a total of ${n\choose p}$ terms.

The
BRST transformation of \eqref{TmABCDdef} is given by
\beq\label{QTm}
QT^m_{A,B,C,D} = k_A^m V_A T_{B,C,D}{} +
\!\!\!\sum_{A=XjY\atop Y=R\shuffle S}\!\!\!
(k_{X}\cdot k_j)\big[
V_{XR}T^m_{jS,B,C,D} - V_{jR}T^m_{XS,B,C,D}\big] + (A\leftrightarrow
B,C,D)\,,
\eeq
for example,
\begin{align}
QT^m_{1,2,3,4} & = k^m_1 V_1 T_{2,3,4} + (1 \leftrightarrow 2,3,4)\,,\\
QT^m_{12,3,4,5} & =  \big[ k_{12}^m V_{12}T_{3,4,5}
+ (12\leftrightarrow 3,4,5)\big]
+ (k_1\cdot k_2)\big(V_1 T^m_{2,3,4,5} - V_2 T^m_{1,3,4,5}\big)\,.\notag
\end{align}
Other building blocks were defined in \cite{oneloopI} to capture the gauge anomaly of the field-theory SYM integrands
that disappear in the $SO(32)$ superstring \cite{green1,green2}.

\medskip\noindent\textbf{Four points}
At genus one, the simplest scattering amplitude with four massless states computed in 1982 by
Green and Schwarz \cite{Green:1981ya} was reproduced in a 2004 calculation
using the minimal pure spinor formalism \cite{MPS}. A salient feature of this calculation is the absence
of OPE singularities among the vertices; the amplitude is completely determined by the pure spinor
zero modes. The result of the correlator in the conventions of \eqref{openagain} is given by
\beq\label{fourptcorr}
{\cal K}_4(\ell) = V_1 T_{2,3,4}\,,
\eeq
and its zero-mode evaluation can be written in terms of the tree-level SYM amplitude $\AYM$ as
follows
\beq\label{looptree}
\langle V_1 T_{2,3,4}\rangle = s_{12}s_{23} A^{\rm SYM}(1,2,3,4)\,.
\eeq
For Neveu-Schwarz external states, the zero-mode evaluation of \eqref{looptree} yields the famous
$t_8$ tensor, $\langle V_1 T_{2,3,4}\rangle = \half t_8(f_1,f_2,f_3,f_4)$ where
\beq\label{t8def}
t_8(f_1,f_2,f_3,f_4) = {\rm tr}(f_1f_2f_3f_4) - {1\over4}{\rm tr}(f_1f_2){\rm tr}(f_3f_4) + {\rm
cyclic}(2,3,4)\,,
\eeq
and ${\rm tr}$ represents a trace over Lorentz indices, for example ${\rm
tr}(f_1,f_2)=f_1^{mn}f_2^{nm}$.

\medskip\noindent\textbf{Five points}
At five points, an analysis of the structure of the superstring correlator arising from the pure spinor
prescription \eqref{presc1loop} reveals that the it is composed of two sectors: one containing a loop momentum contracting
a vectorial combination of superfields and no OPE singularities, and another with no loop momentum and with
singularities as vertex positions collide multiplying a collection of superfields with no free
vector indices. Combining this information with the BRST transformation properties of scalar
\eqref{QTABC} and
vectorial building blocks \eqref{QTm} as well as the monodromy properties of the functions $g^{(n)}(z,\tau)$
and $\ell^m$, yields the proposal for the five-point correlator
\begin{align}
\label{K5}
{\cal K}_5(\ell,z_i)& = V_1 T^m_{2,3,4,5} \ell^m \\
&{}+V_{12} T_{3,4,5} g^{(1)}_{12}  + (2{\leftrightarrow} 3,4,5)\cr
&{}+V_1 T_{23,4,5}g^{(1)}_{23} + (2,3|2,3,4,5)\,.\notag
\end{align}
This is BRST invariant up to total worldsheet derivatives since
\begin{align}\label{Qinv}
Q\cK_5(\ell,z_i)\cI_5(\ell) &= - V_1V_2T_{3,4,5}\big((\ell\cdot k_2)
+s_{21}g^{(1)}_{21}
+s_{23}g^{(1)}_{23}+s_{24}g^{(1)}_{24}+s_{25}g^{(1)}_{25}\big)\cI_5(\ell)\notag\\
&= - V_1V_2T_{3,4,5}{\p\over\p z_2}\cI_5(\ell)\,,
\end{align}
where $\cI_5(\ell)$ is the Koba-Nielsen factor \eqref{IIIKNfactor}.

The correlator \eqref{K5} is also
homology invariant up to BRST-exact terms, around both $A$ and $B$
cycles as a function of $\ell^m$ and $z_i$. In other words, $\cK_5(\ell,z_i)$
is an example of a generalized elliptic integrand \cite{oneloopII}.
To see this note that $\ell^m$ and $g^{(n)}_{ij}$ are single-valued around $A$ cycles
while $\ell^m\to\ell^m -2\pi ik^m_j$ and $g^{(1)}_{ij}\to -2\pi i$ as $z_j$ is moved around the $B$
cycle (with $\tau\to\tau+1$). That is, under the action of the monodromy operator \eqref{Dop} we get
\begin{align}\label{mod1}
D{\cal K}_5(\ell) &= \Omega_1\Big( k_1^m V_1T^m_{2,3,4,5}
+ \big[ V_{12}T_{3,4,5} + 2\leftrightarrow 3,4,5\big]\Big)\\
&{}+\Omega_2\Big( k_2^m V_1T^m_{2,3,4,5} +
V_{21}T_{3,4,5}
+ \big[ V_{1}T_{23,4,5} + 3\leftrightarrow 4,5\big]\Big) +
(2\leftrightarrow 3,4,5)\,,\notag
\end{align}
which can be shown to be BRST-exact \cite{oneloopI} as it is BRST closed and a {\it local}
five-point expression.

\medskip\noindent\textbf{Six points}
Similar considerations of the zero-mode structure from the pure spinor prescription together with
BRST and homology invariance were used to determine the six-point correlator at genus one in
\cite{oneloopIII}. The result can be written as\footnote{As discussed in \cite{oneloopIII}, there is a beautiful Lie-polynomial
compact representation of higher-point genus-one correlators which reveals a common structure with genus zero
correlators and elucidates the combinatorics of \eqref{sixloc}. However, as the notation
requires concepts such as the decreasing Lyndon factorization of words and Lie polynomials
\cite{reutenauer}
we chose to omit it here for brevity.}
\begin{align}
\label{sixloc}
\cK_6(\ell,z_i) &= {1\over2}
V_1 T^{mn}_{2,3,4,5,6}\cZ_{1,2,3,4,5,6}^{mn} \\
&+  V_{12} T^m_{3,4,5,6} \cZ^m_{12,3,4,5,6}
+ (2\leftrightarrow 3,4,5,6)\cr
&+
 V_1 T^m_{23,4,5,6}\cZ^m_{1,23,4,5,6}  + (2,3|2,3,4,5,6) \cr
&+V_{123}T_{4,5,6} \cZ_{123,4,5,6} +V_{132} T_{4,5,6} \cZ_{132,4,5,6}
+ (2,3|2,3,4,5,6) \cr
& + V_1 T_{234,5,6} \cZ_{1,234,5,6}
+ V_1 T_{243,5,6} \cZ_{1,243,5,6}
+ (2,3,4|2,3,4,5,6)\cr
& + \big[\big(V_{12} T_{34,5,6}\cZ_{12,34,5,6} + {\rm cyc}(2,3,4)\big)
 + (2,3,4|2,3,4,5,6)\big]\cr
 &+\big[\bigl(V_{1}T_{2,34,56} \cZ_{1,2,34,56}   + {\rm
 cyc}(3,4,5)\bigr)+ (2\leftrightarrow 3,4,5,6) \big]\,, \notag
\end{align}
where the shorthand for the worldsheet functions are
\begin{align}
\label{newgsAgain}
\cZ_{123,4,5,6}&= g^{(1)}_{12}g^{(1)}_{23} + g^{(2)}_{12} +
g^{(2)}_{23} - g^{(2)}_{13}\,, \\
\cZ_{12,34,5,6}&= g^{(1)}_{12}g^{(1)}_{34}
+ g^{(2)}_{13} + g^{(2)}_{24}
- g^{(2)}_{14} - g^{(2)}_{23}\,,\cr
\cZ^m_{12,3,4,5,6}&= \ell^m g^{(1)}_{12} 
+ (k_2^m - k_1^m)g^{(2)}_{12}
+ \big[ k_3^m (g^{(2)}_{13} - g^{(2)}_{23}) + (3\leftrightarrow
4,5,6)\big]\,,\cr
\cZ^{mn}_{1,2,3,4,5,6}&= \ell^m\ell^n +
\bigl[( k_1^{m}k_2^{n}+k_1^{n}k_2^{m}) g^{(2)}_{12} + (1,2|1,2,3,4,5,6)
\bigr]\,.\notag
\end{align}
After a lenghty calculation, the correlator \eqref{sixloc} was shown to be homology invariant up to vanishing 
BRST-exact terms therefore constituting a six-point example of a generalized elliptic integrand.
The analysis of BRST invariance is more subtle as the six-point open-string correlator at genus one 
is anomalous before summing over the different worldsheet topologies including the M\"obius strip
\cite{green1,green2}. Since gauge invariance is reflected on BRST invariance, to study anomalous
correlators the concept of BRST pseudo invariance was introduced in \cite{partI}. The idea is 
that the BRST variation of pseudo-invariant superfields generate anomalous superfields
\beq\label{anomY}
Y_{A,B,C,D,E} = \half (\l\g^m W_A)(\l\g^n W_B)(\l\g^p W_C)(W_D\g_{mnp}W_E)
\eeq
generalizing the pure-spinor superspace expression
found in the six-point anomaly analysis of \cite{anomaly},
$(\l\g^m W_2)(\l\g^n W_3)(\l\g^p W_4)(W_5\g_{mnp}W_6)$, with parity-odd component expansion
\beq
\langle (\l\g^m W_2)(\l\g^n W_3)(\l\g^p W_4)(W_5\g_{mnp}W_6)\rangle =
-{1\over 16}\epsilon_{10}^{m_2n_2 \ldots m_6n_6}F^2_{m_2n_2} \ldots
F^6_{m_6n_6}.
\eeq
This is captured by the correlator \eqref{sixloc} as its BRST variation, after discarding total
worldsheet derivatives, is given by
\beq\label{Qsix}
Q\cK_6(\ell,z_i)\cI_6(\ell) =-  2\pi i \, V_1 Y_{2,3,4,5,6}  {\p\over\p\tau}\log{\cal I}_6(\ell)\,.
\eeq
Thus, the BRST variation is a boundary term in moduli space \cite{FMS} and vanishes due to
the anomaly cancellation effect of summing over the different worldsheet topologies when
the gauge group is $SO(32)$ \cite{green1,green2}.

\medskip\noindent\textbf{Seven points}
A seven-point open-string correlator at genus one was also obtained in \cite{oneloopIII} and can
be written using various kinds of generalized elliptic integrands $E_{ \ldots}^{ \ldots}$
discussed at length in \cite{oneloopII}
\begin{align}
\label{sevenloc}
\cK_7(\ell,z_i) &= {1\over 6} V_1T^{mnp}_{2,3,\ldots,7}E^{mnp}_{1|2,3,\ldots,7} \\
&+ {1\over 2}V_1T^{mn}_{23,4,5,6,7} E^{mn}_{1|23,4,5,6,7} + (2,3|2,3,4,5,6,7)\cr
& +\big[ V_1T^m_{234,5,6,7} E^m_{1|234,5,6,7} +V_1T^m_{243,5,6,7} E^m_{1|243,5,6,7} \big]+
(2,3,4|2,3,4,5,6,7) \cr
%
&+ \big[V_1 T^{m}_{23,45,6,7} E^{m}_{1|23,45,6,7} + {\rm cyc}(2,3,4)\big]+(6,7|2,3,4,5,6,7)\cr
& +\big[V_1T_{2345,6,7} E_{1|2345,6,7} + {\rm perm}(3,4,5)\big] +(2,3,4,5|2,3,4,5,6,7)\cr
& + \big[V_1T_{234,56,7} E_{1|234,56,7} + V_1T_{243,56,7} E_{1|243,56,7}
+ {\rm cyc}(5,6,7)\big] + (2,3,4|2,3,4,5,6,7)\cr
&+ \big[ V_1T_{23,45,67} E_{1|23,45,67} +{\rm cyc}(4,5,6) \big]+ (3\leftrightarrow 4,5,6,7)\cr
& -  V_1J^m_{2|3,4,5,6,7} E^m_{1|2|3,4,5,6,7} +(2\leftrightarrow3,4,5,6,7) \cr
&- V_1J_{23|4,5,6,7} E_{1|23|4,5,6,7}  + (2,3|2,3,4,5,6,7) \cr
&- \big[ V_1J_{2|34,5,6,7} E_{1|2|34,5,6,7}+{\rm cyc}(2,3,4)\big]+(2,3,4|2,3,4,5,6,7)\,.\notag
\end{align}
This was shown to be BRST (pseudo) invariant and also homology invariant up to BRST-exact terms
and total derivatives in the worldsheet and in moduli space.

\subsubsection{One-loop SYM integrands from the cohomology of pure spinor superspace}

Another application of the pure spinor formalism and related ideas resulted
in expressions for the 1-loop integrands of ten-dimensional
SYM theory \cite{tow1}. The idea is to use the zero-mode structure
suggested by the pure spinor prescription, i.e., after removing
non-zero modes via OPEs leading to multiparticle superfields, to directly propose SYM 1-loop integrands
$A(1,2, \ldots,n|\ell)$ governing the integrated single-trace amplitude via
\beq\label{Aint}
A(1,2,3, \ldots,n) = \int {d^D\ell \over (2\pi)^D} \langle A(1,2,3, \ldots,n| \ell) \rangle \,.
\eeq
More precisely, the 1-loop integrands are expanded in terms of cubic graphs $\Gamma_i$
\beq\label{expAint}
A(1,2,3, \ldots,n| \ell) = \sum_{\Gamma_i}
 {N_i(\ell)\over \prod_k P_{k,i}(\ell)} \,,
\eeq
where the sum is over all 1-loop cubic graphs from boxes to $n$-gons, excluding triangles, bubbles
and tadpoles \cite{notriangle}. Note that the superspace
numerators $N_i(\ell)$ and the propagators $P_{k,i}(\ell)$ not only depend on the external
kinematics but also on the loop momentum~$\ell$.
In proposing the integrand \eqref{expAint}, one respects the supersymmetry constraint that
the numerators of a $p$-gon diagram contain at most $p-4$ powers of $\ell$. Furthermore, it is not
difficult to be convinced that overall BRST invariance of the integrand
can be achieved only if each term of $QN_i(\ell)$ has a factor of $P_{k,i}(\ell)$ with
$k=1,2,\ldots,n$. Schematically,
\beq\label{QNansatz}
QN_i(\ell) = \sum P_{k,i}( \ldots)\,,
\eeq
for some subset of $k$ with the ellipsis representing combinations of (multiparticle) superfields.
Integrands up to six points were found in \cite{tow1} following these lines.

\bigskip
\noindent\textbf{Four point integrand} The integrand of the color-ordered amplitude is expressed in
terms of a single box with a BRST-invariant numerator:
\vspace{-2.5pt}
\beq
\hspace{-40pt}\phantom{.}\raisebox{-18.5pt}
{\includegraphics[width=0.37\textwidth]{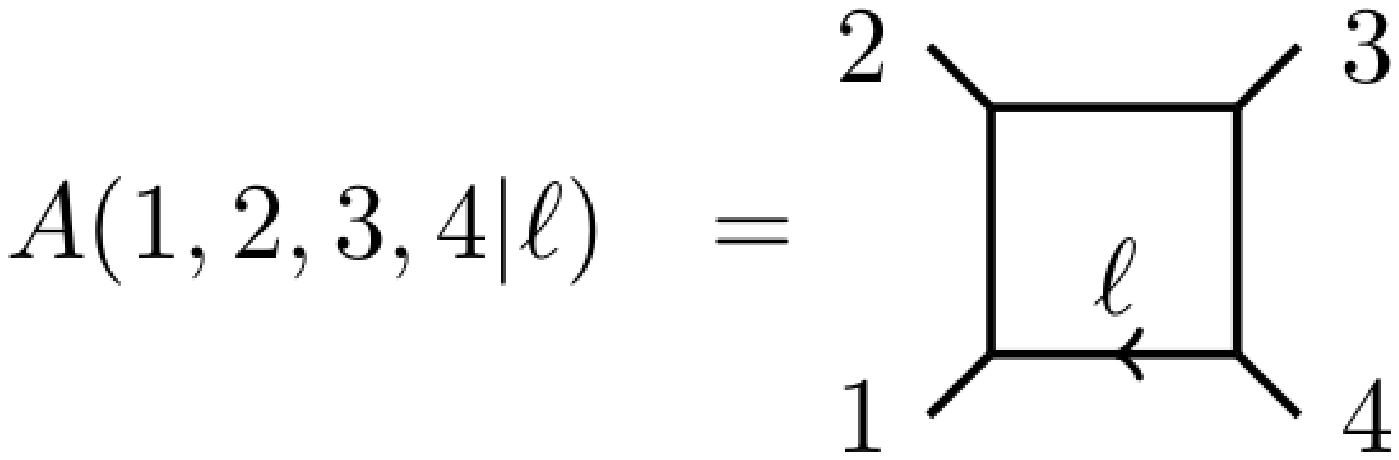}}
\hspace{5pt} = { V_1T_{2,3,4} \over \ell^2 (\ell-k_1)^2 (\ell-k_{12})^2 (\ell-k_{123})^2}\,.
\eeq
This integrand is manifestly BRST invariant using \eqref{QTsone} and agrees with the result obtained
by Green, Brink and Schwarz \cite{Green:1982sw} from the field-theory limit of string theory.

\bigskip
\noindent\textbf{Five point integrand} The integrand of the SYM five-point one-loop amplitude
is expanded in terms of five boxes and one pentagon:
\beq
\hspace{-40pt}\phantom{.}\raisebox{-28.5pt}
{\includegraphics[width=0.81\textwidth]{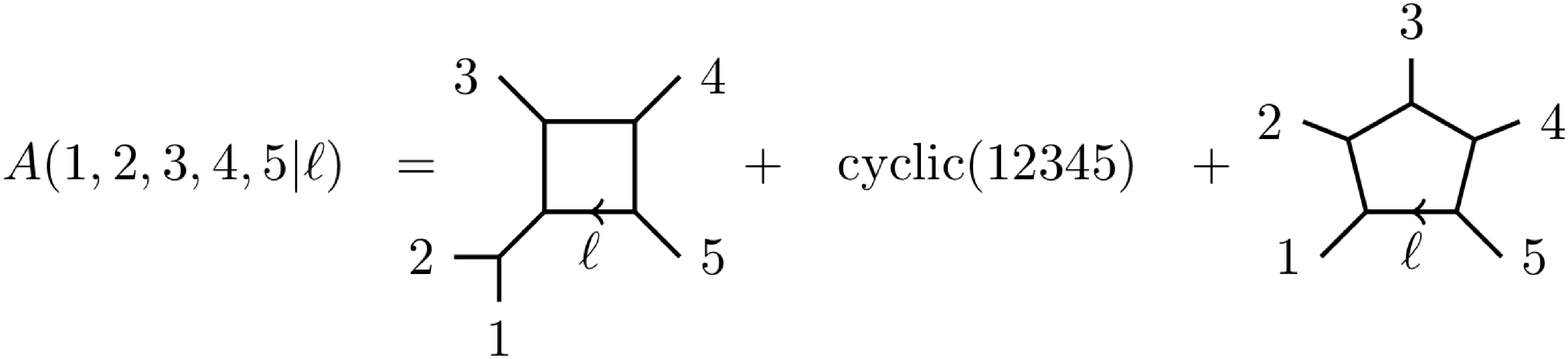}}
\eeq
\[
\hspace{-10pt} {} =  A_{\rm box}(1,2,3,4,5) + A_{\rm pent}(1,2,3,4,5|\ell)
\]
with the corresponding pure spinor superspace expressions given by
\begin{align}
\label{5ptboxes}
A_{\rm box}(1,2,3,4,5) &= {V_{12} T_{3,4,5} \over (k_1+k_2)^2 \ell^2  (\ell-k_{12})^2
(\ell-k_{123})^2 (\ell-k_{1234})^2} \\
&+  {V_{1} T_{23,4,5} \over (k_2+k_3)^2 \ell^2 (\ell-k_1)^2 (\ell-k_{123})^2 (\ell-k_{1234})^2} \cr
&+  {V_{1} T_{2,34,5} \over (k_3+k_4)^2 \ell^2 (\ell-k_1)^2 (\ell-k_{12})^2  (\ell-k_{1234})^2}  \cr
&+  {V_{1} T_{2,3,45} \over (k_4+k_5)^2 \ell^2 (\ell-k_1)^2 (\ell-k_{12})^2 (\ell-k_{123})^2 }\cr
&+  {V_{51} T_{2,3,4} \over (k_1+k_5)^2 (\ell-k_1)^2 (\ell-k_{12})^2 (\ell-k_{123})^2
(\ell-k_{1234})^2} \cr
\label{5ptpent}
A_{\rm pent}(1,2,3,4,5|\ell) &= { N^{(5)}_{1|2,3,4,5}(\ell) \over \ell^2 (\ell-k_1)^2
(\ell-k_{12})^2 (\ell-k_{123})^2 (\ell-k_{1234})^2}
\end{align}
and pentagon numerator
\begin{align}
\label{5ptpentN}
N^{(5)}_{1|2,3,4,5}(\ell) &=
\ell_m  V_1 T^m_{2,3,4,5} + {1\over 2} \big[V_{12} T_{3,4,5} + (2\leftrightarrow 3,4,5) \big]\\
& + {1\over 2} \big[V_{1} T_{23,4,5} + (2,3|2,3,4,5) \big]\,.\notag
\end{align}
To see that this integrand is BRST invariant note that the BRST variation of the local
pentagon numerator satisfies the criterion of canceling pentagon propagators as
\begin{align}
QN^{(5)}_{1|2,3,4,5}(\ell) &=  {1\over 2} V_1 V_2 T_{3,4,5} \big[ (\ell-k_{12})^2- (\ell-k_1)^2
\big]\\
&+{1\over 2} V_1 V_3 T_{2,4,5} \big[ (\ell-k_{123})^2 - (\ell-k_{12})^2 \big]\cr
&+ {1\over 2} V_1 V_4 T_{2,3,5} \big[ (\ell-k_{1234})^2 - (\ell-k_{123})^2\big]\cr
&+{1\over 2} V_1 V_5 T_{2,3,4} \big[\ell^2 - (\ell-k_{1234})^2 \big]\notag
\end{align}
implies that $QA_{\rm pent}(1,2,3,4,5|\ell)$ becomes a sum of boxes of the same type
as contained in $A_{\rm box}(1,2,3,4,5)$. In turn, the BRST variation of the boxes cancels
the external propagators with external momenta $k_i$ in \eqref{5ptboxes} rather than the internal propagators with loop
momentum. Therefore 
the BRST variation of $QA_{\rm box}(1,2,3,4,5)$ is still a sum of boxes,
\begin{align}
QA_{\rm box}(1,2,3,4,5) &= { V_1 V_2 T_{3,4,5} \over 2 \ell^2 (\ell-k_{123})^2 (\ell-k_{1234})^2} \left( {1\over (\ell-k_{12})^2} - {1\over (\ell-k_{1})^2} \right)
\\
&+
{ V_1 V_3 T_{2,4,5} \over 2 \ell^2 (\ell-k_{1})^2 (\ell-k_{1234})^2} \left( {1\over (\ell-k_{123})^2} - {1\over (\ell-k_{12})^2} \right)
\cr
&+
{ V_1 V_4 T_{2,3,5} \over 2 \ell^2 (\ell-k_{1})^2 (\ell-k_{12})^2} \left( {1\over (\ell-k_{1234})^2} - {1\over (\ell-k_{123})^2} \right)
\cr
&+
{ V_1 V_5 T_{2,3,4} \over 2 (\ell-k_1)^2 (\ell-k_{12})^2 (\ell-k_{123})^2} \left( {1\over \ell^2} - {1\over (\ell-k_{1234})^2} \right) \ .
\notag
\end{align}
which ultimately cancels the
variation of the pentagon \eqref{5ptpent}, leading to an overall BRST invariant five-point one-loop integrand. This
example illustrates the mechanism that the BRST variation of a numerator must be engineered to cancel
either internal or external propagators in order to achieve overall BRST invariance.


\bigskip
\noindent\textbf{Six point integrand} The six point integrand is composed of 21 boxes, 6 pentagons
and 1 hexagon 
\beq\label{sixp}
A(1,2,\ldots,6|\ell) = A_{\rm box}(1,2,\ldots,6)
+ A_{\rm pent}(1,2,\ldots,6|\ell) + A_{\rm hex}(1,2,\ldots,6|\ell)\,,
\eeq
whose superspace expressions can be found in \cite{tow1}. A noteworthy feature of the
pure spinor superspace proposal for \eqref{sixp}
is that it leads to an anomalous integrated BRST variation
\beq
\int d^D\ell\, QA(1,2,3,4,5,6|\ell) = -{\pi^5\over 240} V_1Y_{2,3,4,5,6}\,,
\eeq
signaling the well-known fact that the ten-dimensional SYM theory is anomalous at one loop, see \cite{tow1} for
more details.

Note that the six-point one-loop integrand was recently derived in \cite{1loopFTelliot}
in a parameterization satisfying the one-loop color-kinematics duality.

\subsubsection{Genus two}

After the pioneering genus-two calculation with four massless NS states with the RNS formalism in
\cite{2loopDP}, the pure spinor formalism was used in \cite{twoloop,NMPS} to extend the computation to
the supersymmetric graviton multiplet (see also \cite{twolooptwo,mids,2loopH} for explicit
component expansions and the overall normalization factor). For five massless closed-string
states, the supersymmetric amplitudes were computed in the low-energy approximation including their overall
normalization in \cite{2loopH5} and later extended to all orders in $\ap$ in
\cite{2loopI,2loopII}.

The $n$-point amplitude prescription \eqref{presc2loop} gives rise to a chiral amplitude ${\cal
F}_n$ which factorizes into
a Koba-Nielsen factor (in conventions where $s_{ij}=k_i\cdot k_j$)
\beq
\cI_{n} = \exp \Big(  {1\over 4\pi i}\Omega _{IJ} \ell^I  \cdot \ell^J
- \sum_{i=1}^n (\ell^I \cdot k_i) \int^{z_i}_{z_0} \om _I  + \sum_{i<j}^n s_{ij} \ln E(z_i,z_j)\Big)
\label{KNN}
\eeq
and a chiral correlator $\cK_n(\ell^I,z_i)$ carrying the dependence on loop momenta, vertex
operator positions and the polarizations and external momenta of the string states. Since the
vertex positions will be integrated over the Riemann surface one is free to use chiral correlators
which differ by total derivatives as representing the same amplitude under integration by parts
(IBP). For instance, the logarithmic
derivative of the Koba-Nielsen factor is a primary example of an IBP generator:
\beq\label{2IBP}
\p_{z_1}\ln\cI_5 = -(\ell^I\cdot k_1)\om_I(z_1) + s_{12}\eta_{12} + s_{13}\eta_{13}+
s_{14}\eta_{14}+ s_{15}\eta_{15}\,.
\eeq

\bigskip
\noindent\textbf{BRST invariance}
The integration of the zero modes of pure spinor fields together with considerations from group
theory to piece together Lorentz-invariant combinations of superfields initially led to the
introduction of pure spinor superfield building blocks with four external states in
\cite{twolooptwo}
\beq\label{kin2}
T_{1,2|3,4} =
{1\over 64}
(\lambda \gamma_{mnpqr} \lambda) F^{mn}_1 F^{pq}_2 \big[ F^{rs}_3(\lambda \gamma_s W_4)
+ F^{rs}_4(\lambda \gamma_s W_3)\big] + (1,2\leftrightarrow 3,4) \\
\eeq
Considerations involving the
BRST variations of suitable multiparticle superfields allowed an all-multiplicity
generalization of \eqref{kin2} to be found in \cite{tow2}.
Using the language of (minimal) pure spinor superspace, these generalizations have the form \cite{tow2}
\beq\label{s2loop}
T_{A,B|C,D} = {1\over 64}
(\lambda \gamma_{mnpqr} \lambda) F^{mn}_A F^{pq}_B \big[ F^{rs}_C(\lambda \gamma_s W_D)
+ F^{rs}_D(\lambda \gamma_s W_C)\big] + (A,B\leftrightarrow C,D) 
\eeq
In addition, starting from the five-point correlator there are additional Lorentz scalar and tensorial
building blocks
\begin{align}
T^m_{1,2,3|4,5}&=   A_1^m T_{2,3|4,5} + A_2^m T_{1,3|4,5} + A_3^m T_{1,2|4,5} +
W^m_{1,2,3|4,5}\,,\\
T_{1;2|3|4,5} &=
{1\over 2} \Big ((k_1^m{+}k_2^m{-} k_3^m) T^m_{1,2,3|4,5} +
T_{12,3|4,5}+T_{13,2|4,5}+T_{23,1|4,5} \Big )\,,\notag
\end{align}
where $W^m_{1,2,3|4,5}$ is designed in a way as to give the desired BRST variation and symmetry
properties to $T^m_{1,2,3|4,5}$, see below. Its explicit form can be found in \cite{tow2}.
The BRST variation of the four-point building block is given by
\beq\label{Q24}
Q T_{1,2|3,4} = 0\,,
\eeq
while the BRST variation of the five-point building blocks are given by
\begin{align}
\label{Q2s}
Q T_{12,3|4,5} &= s_{12} ( V_1 T_{2,3|4,5} - V_2 T_{1,3|4,5} )\,, \\
Q T_{1;2|3|4,5} &= s_{12}V_1T_{2,3|4,5}\,,\cr
Q T^m_{1,2,3|4,5} &= k_1^m V_1 T_{2,3|4,5} + k_2^m V_2 T_{1,3|4,5} + k_3^m V_3 T_{1,2|4,5} \,.\notag
\end{align}
Furthermore, these building blocks satisfy various crucial identities to capture the correct
features of the
integrand
\begin{gather}
\label{symTs}
T_{A,B|C,D}= T_{B,A|C,D}=T_{C,D|A,B} \ , \quad T_{A,B|C,D}+T_{B,C|A,D}+T_{C,A|B,D}=0 \\
T^m_{1,2,3|4,5}=T^m_{(1,2,3)|(4,5)}\,,\quad
\langle T^m_{1,2,3|4,5} \rangle = \langle T^m_{3,4,5|1,2}+T^m_{2,4,5|1,3}+T^m_{1,4,5|2,3} \rangle
\cr
T_{1;2|3|4,5}= T_{1;2|3|5,4}\,,\quad \langle T_{1;2|3|4,5}+ T_{1;2|4|5,3}+ T_{1;2|5|3,4}\rangle =0\cr
\langle k_3^m (T^m_{1,2,3|4,5}+T^m_{3,4,5|1,2}) - T_{13,2|4,5} -  T_{23,1|4,5} +  T_{34,5|1,2} +
T_{35,4|1,2} \rangle =0\cr
k_1^m T^{m}_{1,2,3|4,5} = T_{2;1|3|4,5}+T_{3;1|2|4,5}\cr
\langle k_5^m T^{m}_{1,2,3|4,5}\rangle = \langle T_{1;5|4|2,3}+T_{2;5|4|1,3}+T_{3;5|4|1,2}\rangle\cr
T_{1;2|3|4,5}-T_{2;1|3|4,5}=T_{12,3|4,5}\cr
\langle T_{5;1|2|3,4}+T_{5;2|1|3,4}+T_{5;3|4|1,2}+T_{5;4|3|1,2}\rangle =0\notag
\end{gather}
where the identities that hold only in the cohomology have been indicated by the pure spinor bracket.

\bigskip
\noindent\textbf{Homology invariance}
The genus two integrands up to five points can be written in terms of the holomorphic
differentials $\om_I$ and loop momenta $\ell^m_I$, the prime form $E(z_i,z_j)$, and single
derivatives of its logarithm $\p_{i} \ln E(z_i, z_j)$. Note that the prime form $E(z,w)$
is holomorphic in $z$ and $w$, odd under $z\leftrightarrow w$, and has a unique simple
zero at $z=w$. Both the loop momentum and the prime form are single valued when $z$ is moved
around $A_I$ cycles, but they have non-trivial
monodromy around a $B_I$ cycle \cite{DHoker:1989cxq}
\begin{align}
\label{monod2}
E(z+B_I ,w) &= - \exp \biggl ( -i \pi \Omega _{II} - 2 \pi i \int ^z _w \! \omega _I \biggr )E(z,w)\\
\p_z \ln E(z+B_I,w) & = \p _z \ln E(z,w) - 2 \pi i \om_I(z) \cr
\p_z \ln E(z,w+B_I) & = \p _z \ln E(z,w) + 2 \pi i \om_I(z)\cr
\ell^m_I &= \ell^m_I - 2\pi ik^m_i\notag
\end{align}
In order to avoid clutter in the formulas below, it is convenient to define the genus-two
propagator as
\beq\label{eta2def}
\eta_{ij} = {\p\over \p z_i}\ln E(z_i,z_j)\,.
\eeq

\noindent\textbf{Basis of holomorphic one-forms} At genus two the holomorphic one-forms $\om_I(z_i)$ are labelled
by $I=1,2$ and modular invariance of the string amplitude suggests that they form $SL(2)$
invariant singlets\footnote{Discussions with Oliver Schlotterer are warmly
acknowledged at this point.}, where $\om_I(z)$ is the $(1)$ of $SL(2)$. At four points, the tensor decomposition
$(1)\otimes(1)\otimes(1)\otimes(1) = 2(0)\oplus 3(2) \oplus(4)$
\cite{LiE} shows that there are two scalars in the decomposition of a four-fold product of $\om_I(z_i)$.
Using the definition
\beq\label{2delt}
\Delta_{ij} = \epsilon^{IJ}\om_I(z_i)\om_J(z_j)
\eeq
the two-dimensional basis of scalars composed of four holomorphic one-forms can be taken in
a cyclic arrangement of labels:
\beq\label{basis4}
\Delta_{12}\Delta_{34},\quad \Delta_{23}\Delta_{41}\,.
\eeq
A third scalar $\Delta_{13}\Delta_{24}$ is not independent as the
antisymmetrization over three
indices vanish
\beq\label{overa}
0=
\epsilon^{I[J}\epsilon^{KL]}
\om_I(z_1)\om_J(z_2)\om_K(z_3)\om_L(z_4)\longrightarrow
\Delta_{13}\Delta_{24} = \Delta_{12}\Delta_{34}-\Delta_{41}\Delta_{23}\,.
\eeq
At five points, the decomposition $(1)\otimes(1)\otimes(1)\otimes(1)\otimes(1) = 5(1)\oplus
4(3) \oplus(5)$ shows that there exists a five-dimensional basis of a five-fold product of
one-forms. These can be taken in a cyclic basis \cite{2loopI}\footnote{An algorithm to arrive at this
basis uses the two identities
$\Delta_{ij}\Delta_{kl}=-\Delta_{il}\Delta_{kj}-\Delta_{ij}\Delta_{lk}$ and
$\om_I(z_i)\Delta_{jk}=-\om_I(z_k)\Delta_{ij}-\om_I(z_j)\Delta_{ki}$ repeatedly until all factors
are in cyclic order. It can be shown that the two identities $\om_I(z_1)\Delta_{24}\Delta_{35}=
-\om_I(z_5)\Delta_{12}\Delta_{34}
-\om_I(z_2)\Delta_{51}\Delta_{34}
+\om_I(z_1)\Delta_{23}\Delta_{45}$ and
$\om_I(z_1)\Delta_{25}\Delta_{34}=
-\om_I(z_5)\Delta_{12}\Delta_{34}
-\om_I(z_2)\Delta_{34}\Delta_{51}$
as well as their permutations are enough to rewrite all products of five one-forms in the
basis \eqref{5ptcycb}.
}
\begin{gather}
\label{5ptcycb}
\om_I(z_1)\Delta_{23}\Delta_{45}\,,\quad
\om_I(z_2)\Delta_{34}\Delta_{51}\,,\quad
\om_I(z_3)\Delta_{45}\Delta_{12}\,,\\
\om_I(z_4)\Delta_{51}\Delta_{23}\,,\quad
\om_I(z_5)\Delta_{12}\Delta_{34}\,.\notag
\end{gather}

\medskip\noindent\textbf{Four points}
The massless four-point chiral integrand was obtained using the minimal pure spinor formalism in
\cite{twoloop} and using the non-minimal formalism in \cite{NMPS}. Luckily, both versions of the 
formalism imply that the chiral integrand is obtained purely from the zero modes of pure spinor
variables. A short analysis of the zero-mode structure of the contributing SYM superfields
together with a group theory analysis of $SO(10)$ scalars in pure spinor superspace using 
a $U(5)$ decomposition of pure spinors implies \cite{twoloop}
\beq\label{4pt2loop}
\cK_4 = \langle T_{1,2|3,4}\rangle\Delta_{41}\Delta_{23} + \langle
T_{4,1|2,3}\rangle\Delta_{12}\Delta_{34}
\eeq
where $T_{i,j|k,l}$
is the kinematic factor \eqref{kin2} in the
minimal pure spinor formalism and $\Delta_{ij}$ is defined in
\eqref{2delt}.
It is easy to see that the chiral correlator \eqref{4pt2loop} is BRST closed using \eqref{Q2s}.
Moreover, it is manifestly single valued as it only depends on the vertex positions  via
$\Delta_{ij}$.

It was shown in \cite{mids} via pure spinor BRST cohomology identities
that
the genus-two kinematic factor \eqref{kin2} is proportional to the four-point tree amplitude
\eqref{SYMtree}:
\beq\label{showM}
\langle T_{1,2|3,4}\rangle = s_{12}^2 s_{23}A(1,2,3,4)\,.
\eeq

\medskip\noindent\textbf{Five points}
Several equivalent expressions for the five-point chiral integrand, emphasizing different properties,
were given in \cite{2loopI}. For instance,
\begin{align}
\label{from15}
\cK_5(\ell^I,z_i) & =  \big[\ell_{m}^I T^m_{1,2,3|4,5} \Delta_{51}\omega_I(z_2)\Delta_{34}
+ {\rm cycl(1,2,3,4,5)}\big] \\
&+ \Big[\eta_{12} \big(  T_{1;2|3|4,5} \Delta_{24}\Delta_{35} +
 T_{1;2|4|3,5} \Delta_{23}\Delta_{45}\big)  + (1,2|1,2,3,4,5)\Big]\notag\\
&+ \Big[\eta_{21}\big(  T_{2;1|3|4,5} \Delta_{14}\Delta_{35} +
 T_{2;1|4|3,5} \Delta_{13}\Delta_{45}\big)
 + (1,2|1,2,3,4,5)\Big]\notag
\end{align}
where the notation $+(i,j|1,2,3,4,5)$ means a sum over all ordered choices of
$i$ and $j$ from the set $\{1,2,3,4,5\}$ for a total of ${5\choose 2}$ terms.

\bigskip
\noindent\textbf{BRST invariance} Using the BRST variation \eqref{Q2s} of the building blocks,
the BRST variation of the chiral correlator \eqref{from15} can be written as
\begin{align}
\label{QK5s}
Q\cK_5(\ell^I,z_i) &= 
V_1T_{5,2|3,4}\Delta_{23}\Delta_{45}(\ell^I\cdot k_1)\om_I(z_1)\\
&+V_1T_{2,3|4,5}\Delta_{12}\Delta_{34}(\ell^I\cdot k_1)\om_I(z_5)\cr
&+V_1T_{2,3|4,5}\Delta_{51}\Delta_{34}(\ell^I\cdot k_1)\om_I(z_2)\cr
&+ V_1\big(T_{2,3|4,5}\Delta_{24}\Delta_{35}  + T_{2,4|3,5}\Delta_{23}\Delta_{45}\big)s_{12}\eta_{12}\cr
&+ V_1\big(T_{3,2|4,5}\Delta_{34}\Delta_{25}  + T_{3,4|2,5}\Delta_{32}\Delta_{45}\big)s_{13}\eta_{13}\cr
&+ V_1\big(T_{4,2|3,5}\Delta_{43}\Delta_{25}  + T_{4,3|2,5}\Delta_{42}\Delta_{35}\big)s_{14}\eta_{14}\cr
&+ V_1\big(T_{5,2|3,4}\Delta_{53}\Delta_{24}  + T_{5,3|2,4}\Delta_{52}\Delta_{34}\big)s_{15}\eta_{15}
+{\rm cyc}(1,2,3,4,5)\notag
\end{align}
To see that the terms proportional to $V_1$ are zero up to a total derivative with respect to $z_1$, after replacing
\beq
\Delta_{12}\Delta_{34}(\ell^I\cdot k_1)\om_I(z_5) =
-\Delta_{51}\Delta_{34}(\ell^I\cdot k_1)\om_I(z_2)
-\Delta_{25}\Delta_{34}(\ell^I\cdot k_1)\om_I(z_1)
\eeq
the terms containing the loop momenta simplify to
\begin{align}
&\big(V_1T_{5,2|3,4}\Delta_{23}\Delta_{45}-V_1T_{2,3|4,5}\Delta_{25}\Delta_{34}\big)(\ell^I\cdot
k_1)\om_I(z_1)\cong \\
&\big(V_1T_{5,2|3,4}\Delta_{23}\Delta_{45}-V_1T_{2,3|4,5}\Delta_{25}\Delta_{34}\big)
(s_{12}\eta_{12}+s_{13}\eta_{13}+s_{14}\eta_{14}+s_{15}\eta_{15})\notag
\end{align}
where the IBP relation \eqref{2IBP} $-(\ell^I\cdot k_1)\om_I(z_1)
+ s_{12}\eta_{12} + s_{13}\eta_{13} + s_{14}\eta_{14} + s_{15}\eta_{15}\cong0$
has been used. Plugging this into
\eqref{QK5s}, the terms containing $s_{12}\eta_{12}$ become
\beq
s_{12}\eta_{12}V_1\Big(
T_{2,3|4,5}(\Delta_{24}\Delta_{35}-\Delta_{25}\Delta_{34})
+(T_{2,4|3,5}
+T_{2,5|3,4})\Delta_{23}\Delta_{45}\Big) = 0
\eeq
where we used the kinematic Jacobi identity $T_{2,4|3,5}+T_{2,5|3,4}=-T_{2,3|4,5}$ as well as
the worldsheet Jacobi identity
$\Delta_{24}\Delta_{35}-\Delta_{25}\Delta_{34}-\Delta_{23}\Delta_{45}=0$. The analysis of the other terms
$s_{1j}\eta_{1j}$ for $j=3,4,5$ is similar and the vanishing of the full BRST variation
\eqref{QK5s} follows from the cyclic
permutations.

\bigskip
\noindent\textbf{Homology invariance} Using the monodromies in \eqref{monod2} one can show that the chiral correlator \eqref{from15}
is single-valued as a function of $\ell_I$ and $z_i$. For instance, moving $z_1$ around
the $B_I$ cycle and writing the result in terms of the cyclic basis \eqref{5ptcycb}
implies that \eqref{from15} is single valued around $z_1$ provided
\begin{align}
\langle k^m_1 T^m_{3,4,5|1,2}
          - T_{3;1|2|4,5}
          - T_{4;1|2|3,5}
          - T_{5;1|2|3,4}\rangle &= 0\\
\langle k^m_1  T^m_{1,2,3|4,5}
          + T_{1;4|5|2,3}
          + T_{1;5|4|2,3}
          + T_{12,3|4,5}
          + T_{13,2|4,5}\rangle &= 0\notag
\end{align}
which can be verified to be true using the identities in \eqref{symTs}. Alternatively, their
validity also follows from the fact that these are BRST-closed linear combinations of local building blocks 
and that the five-point local cohomology is empty.

\subsubsection{Genus three}

\medskip\noindent\textbf{Four points}
The chiral correlator for four external  massless states was determined in \cite{3loopH} up to
terms that have no singularities on the worldsheet\footnote{We note a recent \cite{Geyer:2021oox}
conjecture for the full bosonic correlator obtained from matching its
field-theory limit with the ${\cal N}=8$ integrand in a BCJ parameterization.}. It can be written as
\begin{align}\label{K3loop}
\cK_4(\ell) &= T^m_{1,4|2|3} \ell^I_m w^I_1\Delta_{234}
+ T^m_{2,4|1|3} \ell^I_m w^I_2\Delta_{134}
+ T^m_{3,4|1|2} \ell^I_m w^I_3\Delta_{124}\\
&+  T_{12|3|4}\Delta_{234}\,\eta_{12}
+ T_{13|2|4}\Delta_{324}\,\eta_{13}
+ T_{14|2|3}\Delta_{423}\,\eta_{14} \cr
&+ T_{23|1|4}\Delta_{314}\,\eta_{23}
+ T_{24|1|3}\Delta_{413}\,\eta_{24}
+ T_{34|1|2}\Delta_{412}\,\eta_{34}\notag
\end{align}
where $\Delta_{ijk} = \epsilon^{IJK}\om_I(z_i)\om_J(z_j)\om_K(z_k)$ for $I,J,K=1,2,3$ and
$\eta_{ij}$ is a worldsheet function depending on the genus-three prime form $E(z_i,z_j)$
\beq\label{etadef}
\eta_{ij} = {\p\over \p z_i}\ln E(z_i,z_j).
\eeq
The building blocks above depend on non-minimal pure spinor fields. The vectorial building block
is constructed as follows
\beq\label{Tmdef}
T^m_{1,2|3|4} = L_{1342}^m + L_{2341}^m + {5\over 2} S_{1234}^m\,
\eeq
where $S^m_{1234} = S^{(1)\,m}_{1234} + S^{(2)\,m}_{1234} - S^{(2)\,m}_{1243}$
and
\begin{align}\label{Sms}
S^{(1)\,m}_{1234} &= 2\,(\lb\g^m \g^{a_1}\l)(\lb\g_{m_1 n_1 p_1} r) (\lb\g_{m_2 n_2 p_2} r)
(\lb\g_{m_3 n_3 p_3} r)(\lb\g_{m_4 n_4 p_4} r)(\lb\g_{m_5 n_5 p_5} r)\cr
&\times (\l\g^{a_2m_1n_1p_1m_3}\l)(\l\g^{a_3m_2n_2p_2m_5}\l)(\l\g^{n_3m_4n_4p_4n_5}\l)\cr
&\times (W^1\g^{a_1a_2a_3}W^2)(\l\g^{p_3}W^3)(\l\g^{p_5}W^4)\\
S^{(2)\,m}_{1234} &= 96\,(\lb\g^m \g^{m_3}\l)(\lb\g_{m_1 n_1 p_1} r) (\lb\g_{m_2 n_2 p_2} r) (\lb\g_{m_3 n_3 p_3} r)(\lb\g_{m_4 n_4 p_4} r)(\lb\g_{m_5 n_5 p_5} r)\cr
&\times (\l\g^{m_1m_2n_2p_2m_5}\l)(\l\g^{n_3m_4n_4p_4n_5}\l)\cr
&\times (\l\g^{n_1}W^1)(\l\g^{p_1}W^2)(\l\g^{p_3}W^3)(\l\g^{p_5}W^4)\cr
L_{ijkl}^m &=(\lb\g^{abc}r)(\lb\g^{def}r)(\lb\g^{ghi}r)(\lb\g^{mnp}r)(\lb\g^{qrs}r)(\lb\g^{tuv}r)\cr
 &\quad\times(\l\g^{adefm}\l)(\l\g^{bghit}\l)(\l\g^{uqrsn}\l)(\l\g^c W_i)(\l\g^p W_j)(\l\g^v
 W_k)A^m_l.\notag
\end{align}
The scalar building block is given by
\begin{align}\label{Tij3}
T_{ij|k|l} &=
(\lb\g^{abc}r)(\lb\g^{def}r)(\lb\g^{ghi}r)(\lb\g^{mnp}r)(\lb\g^{qrs}r)(\lb\g^{tuv}r)\\
 &\quad\times(\l\g^{adefm}\l)(\l\g^{bghit}\l)(\l\g^{uqrsn}\l)(\l\g^c W_{ij})(\l\g^p W_k)(\l\g^v
 W_l)\,.\notag
\end{align}
The presence of the non-minimal fields $\lb_\a,r_\b$ in these building blocks leads to technical
challenges that do not exist when dealing with ``minimal'' pure spinor superspace expressions. The
$r_\b$ fields can be straightforwardly converted into superspace derivatives $D_\b$ but the handling
of $\lb_\a$ is not so immediate. But
luckily, as proven in the appendix of \cite{3loopH}, there exists a procedure to convert an
arbitrary non-minimal pure spinor superspace expression containing $\lb^n\l^{n+3}$ pure spinors with $n\ge1$
into an expression in which the $\lb_\a$ are contracted with $\l^\a$ yielding ``minimal'' pure
spinor superspace expressions with
$(\l\lb)^n \l^3$. As the $(\l\lb)^n$ factor only affects the normalization of the zero-mode
integration, one can consider these non-minimal pure spinor superspace expressions more or less
in the same footing as their minimal counterparts. It is worth mentioning that there is a proposal
for these building blocks directly in minimal pure spinor superspace using higher-mass SYM
superfields as \cite{genseries}
\begin{align}
T_{12,3,4} &\equiv \langle (\l\g_m W_{12}^n)(\l\g_n W_{[3}^p)(\l\g_p W_{4]}^m)\rangle
\notag \\
T^m_{1234} &\equiv \langle A^m_{(1} T_{2),3,4}^{\phantom m}
+(\l\g^m W_{(1}^{\phantom m}) L^{\phantom m}_{2),3,4}\rangle \label{new3loop} \\
L_{2,3,4} &\equiv \tfrac{1}{3}(\l\g^n W_{[2}^q)(\l\g^q W_{3}^p)F_{4]}^{np}\rangle \,,
\notag
\end{align}
where $W_P^{m\a}$ represents a local superfield of higher-mass dimension as defined in
\cite{genseries}; when $P=i$ is a single letter it reduces to $W_i^{m\a} = k_i^m W^\a_i$ but
when $P$ is a word there are non-trivial contact-term corrections. The
component expansion of these building blocks is not exactly the same as their
non-minimal counterparts but they yield the same $D^6 R^4$ components as
discussed below.

After approximating the Koba-Nielsen factor to one in the low-energy limit and
integrating over the volume of moduli space, the holomorphic square of the integrand
\beq\label{d6r4}
{|T_{12,3,4}|^2\over s_{12}} + |T^m_{1234}|^2 + (1,2|1,2,3,4)
\eeq
is proportional
to the $D^6 R^4$ interaction of type II when
expanded in bosonic components, regardless of the minimal vs non-minimal representations of the building
blocks. One can show that the low-energy contribution \eqref{d6r4}
is BRST closed, but not the chiral correlator \eqref{K3loop}. A BRST-closed and single-valued
chiral correlator to all orders in $\ap$ has since been found \cite{wip}.

It is worth noting that the computations of \cite{3loopH} were done keeping track of the absolute
normalizations coming from the pure spinor prescription with the integration formulas from
\cite{humberto}. As will be reviewed below, these calculations matched the predictions to the
$D^6R^4$ type IIB interaction arising
from the S-duality considerations of Green and Vanhove \cite{GreenVanhove3loops}.

\subsection{Verifying S-duality conjectures}

The scattering amplitudes computed with pure spinor formalism have provided
an independent check on the S-duality predictions of type IIB interactions.

\subsubsection{S-duality and four-point amplitudes}

On the one hand, the $SL(2,\mathbb{Z})$-duality prediction for the perturbative four-graviton
type IIB effective action in the string frame is given by
\cite{GreenGutperle,GreenGutperleVanhove,GreenKwonVanhove2loops,GreenVanhove3loops}
\begin{align}\label{Action}
S_{\rm IIB}^{4{\rm pt}} = \int d^{10}x\sqrt{-g}\,\big[& R^4(2\zeta_3 e^{-2\phi} + 4 \zeta_2)
+ D^4R^4(2\zeta_5 e^{-2\phi} + {8\over 3}\zeta_4e^{2\phi}) \\
&+ D^6R^4(4\zeta_3^2 e^{-2\phi} + 8\zeta_2\zeta_3 + {48\over 5}\zeta_2^2 e^{2\phi}
+ {8\over 9}\zeta_6 e^{4\phi}) + \cdots \big],\notag
\end{align}
where the shorthands $R^4$, $D^4R^4$ and $D^6R^4$ denote contractions
of covariant derivatives $D$
and Riemann curvature tensors $R$ whose precise structure does not affect the analysis.
Factors of $e^{(2g-2)\phi}$ are associated with the genus-$g$ order in string perturbation theory.
The key idea of the S-duality analysis was to associate the coefficients of the $R^4$ interaction
with the zero-modes of non-holomorphic Eisenstein series $E_{3/2}(\Phi,\bar\Phi)$ and those of $D^4R^4$ 
with $E_{5/2}(\Phi,\bar\Phi)$, where
\begin{align}\label{eisth}
E_{3/2}(\Phi,\bar \Phi) &= 2\zeta_3 e^{-3\phi/2} + 4 \zeta_2 e^{\phi/2} + \cdots\\
E_{5/2}(\Phi,\bar \Phi) &= 2\zeta_5 e^{-5\phi/2} + {8\over
3}\zeta_4e^{3\phi/2} + \cdots\notag
\end{align}
where $\Phi$ depends on the complex axio-dilaton field $\Phi= C_0+ie^{-\phi}$.

On the other hand, the $\ap$ expansion of perturbative string scattering amplitude calculations performed with
the non-minimal pure spinor formalism with the absolute normalization techniques from
\cite{humberto,2loopH,3loopH} for the four-point massless closed string states are given by
\begin{align}
\label{ampsH}
M^{(0)}_4 &= (2\pi)^{10}\d^{10}(k)\halfap3 \kappa^4 e^{-2\l}\,2\pi K\tilde K\big(
{3\over \s_3} + 2\zeta_3 + \zeta_5 \s_2 + {2\over 3}\zeta_3^2 \s_3 + \cdots
\big) \notag\\
M^{(1)}_4 &= (2\pi)^{10}\d^{10}(k)\halfap3 \k^4 \Big({1 \over 2^4\cdot 3 \pi}\Big)K\tilde K
\big(1 + {\zeta_3\over 3}\sigma_3 + \cdots \big) \notag\\
M_4^{(2)} &= (2\pi)^{10}\d^{10}(k)\halfap3\k^4 e^{2\l}\Big( {1\over 2^{10}\cdot 3^3\cdot 5\pi^3}\Big)K\tilde K\,
\big(\sigma_2 +3\sigma_3 + \cdots  \big) \cr
M_4^{(3)} &= (2\pi)^{10}\d^{10}(k)\halfap3\k^4 e^{4\l}\Big({1\over 2^{15}\cdot 3^6\cdot 5\cdot7\pi^{5}}\Big)K\tilde K\,
\big(\sigma_3 +\cdots \big)
\end{align}
where 
\beq
\s_n = \halfap{n}( s_{12}^n+s_{13}^n+s_{14}^n)
\eeq
are dimensionless symmetric polynomials of Mandelstam invariants $s_{ij}=(k_i\cdot k_j)$, $e^{(2g-2)\l}$ is
the string genus-$g$ coupling constant,
$K$ is the supersymmetric kinematic factor\footnote{For bosonic external states, note that $-2^3K^{\rm here}= K^{\rm 0503}$ from
\cite{DHokerS} and that $K^{\rm here}\tilde K^{\rm here} = \cK_4^{(0)}$ from \cite{2loopH5}. In addition, the amplitudes
in \eqref{ampsH} were computed using the tree-level normalization convention encoded by $R^2=\pi^5/2^5$ used
in \cite{2loopH5} where $R$ is a normalization parameter appearing in the zero-mode measures $[dr]$
and $[ds^I]$. In \cite{3loopH} the normalization $R^2=\sqrt{2}/(2^{16}\pi)$ was chosen, such that 
the genus $g$ amplitudes $A_g^{\rm 1308}$ of \cite{3loopH} are related by $x^{1-g} A_g^{\rm 1308}=
M_g^{\rm 1504}$ to the amplitudes
$M_g^{\rm 1504}$ of \cite{2loopH5} with $x=\sqrt{2}2^{10}\pi^6$ after considering that 
$K^{\rm 1308}\bar K^{\rm 1308}=2^6\cK_4^{(0)}$.}
\beq\label{Kdef}
K = s_{12}s_{23}\AYM(1,2,3,4)
\eeq
and $\k$ is the normalization of the vertex operators fixed to $\k^2 = e^{2\l}\pi/\ap^2$ by unitarity
\cite{2loopH5}.

It is easy to see the one-to-one correspondence of the genus- and $\alpha'$-orders in the
amplitudes \eqref{ampsH} with the curvature
couplings in the action \eqref{Action}
\beq\label{corresp}
e^{(2g-2)\phi}D^{2k}R^4 \leftrightarrow  e^{(2g-2)\l}K\tilde K \s_k\,.
\eeq
Therefore, matching the ratio $\hbox{(genus one)}/\hbox{(genus zero)}$
of the $R^4$ interactions in the effective action with the corresponding ratio
of the amplitudes
\beq
{4\zeta_2 R^4\over 2\zeta_3 e^{-2\phi}R^4} =
{e^{2\phi}\pi^2\over 3\zeta_3}\,,\qquad
{(K\tilde K)/(2^4 3\pi)\over (K\tilde K)4\pi\zeta_3 e^{-2\l}} = {e^{2\l}\over 2^6\,3\pi^2\zeta_3}
\eeq
relates the coupling constants $e^\phi$ and $e^\l$,
\beq\label{Rfourratio}
 e^{2\l} = 2^6 \pi^4\,e^{2\phi} \ .
\eeq
\noindent\textbf{$D^4R^4$ interaction at genus two}
One can now compare the predicted $D^4R^4$ interaction terms from the type IIB effective action \eqref{Action}
with the first principles string calculations. Taking the genus-two/genus-zero ratio of the $K\tilde K \s_2$ term from
the amplitudes gives
\beq
{e^{4\l}\over 2^{11}3^3 5\pi^4\zeta_5} = {2\pi^4 e^{4\phi}\over 3^3 5\zeta_5}
\eeq
where we used \eqref{Rfourratio} in the right-hand side. This is the same ratio
of the $D^4R^4$ terms in \eqref{Action} at the corresponding loop order:
$8\zeta_4e^{4\phi}/(6\zeta_5)=2\pi^4 e^{4\phi}/(3^3 5\zeta_5)$ as $\zeta_4 = \pi^4/(2\cdot 3^2\cdot 5)$.

\bigskip
\noindent\textbf{$D^6R^4$ interaction at genus three} Similarly, the ratio
$\hbox{(genus three)}/\hbox{(genus one)}$ correction $K\tilde K\s_3$
matches perfectly with the S-duality result in \eqref{Action}. The ratio of the amplitudes
is given by $e^{4\l}/(2^{11}\cdot3^4\cdot 5\cdot 7\pi^4\zeta_3)$ while in the effective action
it is given by $\zeta_6e^{4\phi}/(9\zeta_2\zeta_3)$, and these two numbers match after using the conversion \eqref{Rfourratio}
and $\zeta_6 = \pi^6/(3^3\cdot 5\cdot 7)$.

\bigskip
\noindent\textbf{$D^6R^4$ interaction at genus two} The coefficient of $K\tilde K\s_3$ at
genus-two was computed in \cite{ZK2loop} and allowed the comparison between the string scattering
amplitude result at genus two with the S-duality prediction in the action \eqref{Action}. The
$\hbox{(genus two)}/\hbox{(genus one)}$ ratio of the correction $K\tilde K\s_3$ is given by
$e^{2\l}/(2^6 5\pi^2\zeta_3)$ which matches the S-duality $\hbox{(genus two)}/\hbox{(genus
one)}$ ratio of the $D^6R^4$ interaction, given by  $6\zeta_2 e^{2\phi}/(5\zeta_3)$ after using
the conversion \eqref{Rfourratio} and $\zeta_2 = \pi^2/6$.

\subsubsection{S-duality and five-point amplitudes}

The pure spinor formalism also allowed to check the S-duality proposals for five graviton
interactions as well as four gravitons and one dilaton. For five gravitons the S-duality
effective action contains the same ratios appearing in the four graviton action \eqref{Action};
the extension is straightforward with four-curvature corrections $D^{2k}R^4$ followed
by a tail of operators $D^{2(k-p)}R^{4+p}$, although there might be novel $D^{2k}R^{\ge5}$ couplings
without a four-field counterpart such as the $D^6R^5$ interaction at genus one \cite{GMS}.
These S-duality tails such as $(D^4R^4 + D^2R^5)$
are confirmed by the data of the genus-$g$ amplitudes $M_5^{(g)}$ at five points\footnote{To avoid cluttering, we omit the universal factor
of $(2\pi)^{10}\d^{10}(k)$ from the right-hand side of \eqref{FiveAmps}.} \cite{2loopH5}
\begin{align}
\label{FiveAmps}
M^{(0)}_5 &= \halfap{}\; \kappa^5 e^{-2\l}(2\pi)^2 \cK_5^{(0)} \\
M^{(1)}_5 \big|^{\ap^4}_{{\rm IIB}}&= \halfap{} {\k^5 \over  2^{4} 3 }\, \cK_5^{(0)}\Big|_{\zeta_3} \times
\begin{cases}\ \ \, 1 \,  \ : \hbox{five gravitons} \cr
-{1\over3} \ : \hbox{four  gravitons, one dilaton}
\end{cases}\cr
M^{(2)}_5 \big|^{\ap^6}_{{\rm IIB}}&= \halfap{}
{\kappa^5 e^{2\l} \over 2^{9} \, 3^{3}\,5\,\pi^2} \cK_5^{(0)}\Big|_{\zeta_5}\times
\begin{cases}
\ \ \, 1 \,  \ : \hbox{five gravitons} \cr
-{3\over 5} \ : \hbox{four gravitons, one dilaton}
\end{cases}\notag
\end{align}
where the tree-level factor $\cK_5^{(0)}$ is given by
\beq\label{fivetree}
\cK_5^{(0)} = \tilde A^T_{54} \! \cdot \! S_0 \! \cdot \! \big[ 1 + 2\zeta_3\halfap3 M_3 + 2\zeta_5\halfap5 M_5 +2
\zeta_3^2 \halfap6 M_3^2 +  {\cal O}(\ap^7)\big] \! \cdot \! A_{45} \, ,
\eeq
where $\tilde A^T_{54}$ and $A_{45}$ are two-component vectors of SYM tree-amplitudes
\beq\label{inKLT}
 \tilde A_{54}\equiv \begin{pmatrix}\tilde A^{\rm YM}(1,2,3,5,4)\cr
 \tilde A^{\rm YM}(1,3,2,5,4)\end{pmatrix}\,,\quad
  A_{45} \equiv \begin{pmatrix} A^{\rm YM}(1,2,3,4,5)\cr A^{\rm YM}(1,3,2,4,5)\end{pmatrix} \, ,
\eeq
$S_0$ denotes the KLT matrix and
the $2\times 2$ matrices $M_{2n+1}$ were introduced in \cite{motivic}.

Since the calculations in the pure spinor formalism are supersymmetric and done exploiting pure
spinor superspace, the scattering of any state in the graviton supermultiplet can be systematically
obtained once the superspace expression is calculated. As can be seen in \eqref{FiveAmps},
the ratios of the string amplitudes
depend on the R-symmetry charges of the external type IIB states, as trading one graviton for a
dilaton gives the additional factors of $-{1\over 3}$ or
$-{3\over 5}$.

These numbers can be explained by the following argument \cite{sixteen}:
scattering processes which violate the R-symmetry of type IIB supergravity
are associated with operators which transform with modular weight under S-duality, therefore
by modular invariance of the type IIB action, 
they must be accompanied by modular forms of opposite weights to preserve the modular invariance
of the type IIB effective action. These modular forms can be generated as $DE_s$ where $D$ is the
modular covariant derivative such that $De^{q\phi} = q \cdot e^{q\phi}$
and $E_s$ a Eisenstein series. For example,
\begin{align}
DE_{3/2}(\Phi,\bar \Phi) &=  \Big(-{3\over 2} \Big)2\zeta_3 e^{-3\phi/2} +\Big({1\over 2} \Big)
4 \zeta_2 e^{\phi/2}+\cdots \\
D E_{5/2}(\Phi,\bar \Phi) &=
\Big(-{5\over 2} \Big)2\zeta_5 e^{-5\phi/2} +\Big({3\over 2} \Big) {8\over
3}\zeta_4e^{3\phi/2} +\cdots \,.\notag
\end{align}
Thus the ratio between tree-level and higher-genus contributions is
deformed by $-{1\over 3}$ and $-{3\over 5}$ in cases of
$E_{3/2}$ and $E_{5/2}$, suggesting that the type IIB effective action contains the terms
\beq
\int d^{10}x\sqrt{-g}\big[ \phi R^4(-3\zeta_3e^{-2\phi}+ 2\zeta_2)
+ \phi D^4R^4(-5\zeta_5e^{-2\phi}+ 4\zeta_4e^{2\phi})\big]
\eeq
in the string frame\footnote{The term $\phi R^4$ is multiplied by $e^{-\phi/2}$ and
$\phi D^4R^4$ by $e^{\phi/2}$ in going to the string frame.} and
explaining the relative coefficients in the scattering amplitudes \eqref{FiveAmps}.

\vskip0.3in
\textbf{Acknowledgements:} CRM is supported by a University Research Fellowship from the Royal Society.
NB would like to thank CNPq grant 311434/2020-7 and FAPESP grants
2016/01343-7, 2021/14335-0, 2019/21281-4 and 2019/24277-8 for partial
financial support.

\end{document}